\documentclass[12pt,preprint]{aastex}

\newcommand{\kms}{km s$^{-1}$}
\newcommand{\kmsm}{\mathrm{km\;s^{-1}}}
\newcommand{\smpy}{M$_\sun$ yr$^{-1}$}
\newcommand{\smpym}{M_\sun\;\mathrm{yr}^{-1}}

\shorttitle{Galactic Winds in ULIGs}
\shortauthors{Rupke, Veilleux, \& Sanders}

\begin{document}

\title{Keck Absorption-Line Spectroscopy of Galactic Winds in Ultraluminous Infrared Galaxies}
\author{David S. Rupke\altaffilmark{1,2}, Sylvain Veilleux\altaffilmark{1,2,3}, and D. B. Sanders\altaffilmark{2,4}}

\altaffiltext{1}{Department of Astronomy, University of Maryland, College Park, MD 20742; drupke@astro.umd.edu, veilleux@astro.umd.edu}
\altaffiltext{2}{Guest observers at the W.M. Keck Observatory, which is operated as a scientific partnership among the California Institute of Technology, the University of California and the National Aeronautics and Space Administration. The Observatory was made possible by the generous financial support of the W.M. Keck Foundation.}
\altaffiltext{3}{Cottrell Scholar of Research Corporation}
\altaffiltext{4}{Institute for Astronomy, University of Hawaii, 2680 Woodlawn Drive, Honolulu, HI 96822; sanders@ifa.hawaii.edu}

\begin{abstract}
In this paper, we present moderately-high resolution ($\sim 65\;\kmsm$) spectroscopy, acquired with ESI on Keck II, of 11 ultraluminous infrared galaxies at $z < 0.3$ from the IRAS 1 Jy sample.  The targets were chosen as good candidates to host galaxy-scale outflows, and most have infrared luminosities dominated by star formation.  We use a $\chi^2$ minimization to fit one- to three-component profiles to the \ion{Na}{1} D interstellar absorption doublet in each object.  Assuming that gas blueshifted by more than 70 \kms\ relative to the systemic velocity of the host is outflowing, we detect outflows in $73\%$ of these objects.  We adopt a simple model of a mass-conserving free wind to infer mass outflow rates in the range $\dot{M}_{\mathrm{tot}}$(H)$\;= 13-133\;\smpym$ for galaxies hosting a wind.  These values of $\dot{M}_{\mathrm{tot}}$, normalized to the corresponding global star formation rates inferred from infrared luminosities, are in the range $\eta \equiv \dot{M}_{\mathrm{tot}} / \mathrm{SFR} = 0.1-0.7$.  This is on average a factor of only 10 less than $\eta$ from recent measurements of nearby dwarfs, edge-on spirals, and  lower-luminosity infrared galaxies.  Within our sample, we conclude that $\eta$ has no dependence on the mass of the host (parameterized by host galaxy kinematics and absolute $R$- and $K^{\prime}$-band magnitudes).  We also attempt to estimate the average escape fraction $\langle f_\mathrm{esc} \rangle \equiv \sum \dot{M}_\mathrm{esc}^i / \sum \dot{M}_{\mathrm{tot}}^i$ and ``ejection efficiency'' $\langle\delta\rangle \equiv \sum \dot{M}_\mathrm{esc}^i / \sum \mathrm{SFR}^i$ for our sample, which we find to be $\sim 0.4-0.5$ and $\sim 0.1$, respectively.  The complex absorption-line properties of Mrk 231, an ultraluminous infrared galaxy which is optically classified as a Seyfert 1, are discussed separately in an appendix.

\end{abstract}


\section{INTRODUCTION} \label{intro}

Large-scale galactic outflows, energized by stellar winds and supernovae ejecta or a central AGN, are ubiquitous in the local universe and at high redshift \citep*{pet01,fbb02}.  Theoretical work, both analytical and numerical, suggests that these outflows may be important in galaxy formation \citep*[e.g.][]{ds86,sfb00,sb01,std01}.  They also play a role in galactic evolutionary processes, especially the regulation of star formation by mechanical feedback \citep*[e.g.][]{ocm01} and the enrichment of galactic halos.  The expulsion of metals from galaxies by galactic winds may be able to reproduce the color-magnitude (or mass-metallicity) relation of ellipticals, since more massive galaxies, with larger gravitational potentials, could retain mass and metals more easily in this scenario \citep{larson74,vader86,franx90,kc98}.  Besides their impact on individual galaxies, galactic winds may contribute significantly to the chemical and thermal evolution of the universe by enriching and heating the intergalactic medium \citep*[e.g.][]{nath97,aguirre01,mfr01}, which is known to contain substantial amounts of metals even at high redshift and low density \citep[e.g.][]{cs98}.  Winds are also a possible mechanism for the preheating and enrichment of the intracluster medium in groups and clusters of galaxies \citep*[e.g.][]{dfj91,pcn99}.  Finally, these outflows could be responsible for the damped Ly$\alpha$ or strong \ion{Mg}{2} absorption systems observed in quasar spectra \citep[e.g.][]{efsg00,chen01,bond01}, and may contribute to reionization by opening a path for Lyman continuum photons to escape star-forming regions \citep*{mhr99,dsf99,spa01,pet01}.

The observational dataset on galactic-scale outflows is slim, however, apart from studies of the local universe ($z<0.1$) and high-redshift galaxies [Lyman-break galaxies at $z\sim3$ \citep{pet00,pet01} and a few gravitationally-lensed objects \citep{fbb02}].  In this paper, we present and discuss the results of a moderately-high resolution ($R\sim4600$, or $\Delta v \sim 65\;\kmsm$) spectroscopic study of winds in ultraluminous infrared galaxies (ULIGs) with redshifts of $0.04<z<0.27$.  ULIGs, which by definition have $L_{\mathrm{IR}}>10^{12}L_{\sun}$\footnote{$L_{\mathrm{IR}}=L(8-1000)\;\micron$, computed using the single-temperature dust-emissivity fit to all four IRAS flux bands given in \citet{per87} \citep[see also][]{sm96}, $H_0 = 75\;\kmsm\;\mathrm{Mpc^{-1}}$, and $q_0 = 0$.}, contain massive starbursts and/or AGN \citep*{vks99b,lvg99}, and have been associated with large-scale outflows at low redshift \citep*[e.g.][]{ham90,kvs98}.  ULIGs, especially those whose energy output is dominated by stars, are important sites of obscured star formation at low $z$ \citep{sm96}.  There are suggestions from infrared and submillimeter counts that there is a strong increase in the number density of ULIGs with $z$ (IRAS: \citealt{ks98}; ISO: \citealt{kawara98,puget99,matsuhara00,efsa00,serjeant01}; SCUBA: \citealt*{sib97}; \citealt{hughes98,blain99,eales99}; \citealt*{bcs99}), implying that ULIGs contain a substantial fraction of high-$z$ star formation (although this is not necessarily true; see \citealt*{tbg99}).  ULIGs may also be a highly reddened and luminous subset of the UV-selected Lyman-break galaxies \citep[e.g.][]{sand99, tbg99}, which typically produce outflows \citep{pet01}.

Absorption-line spectroscopy has proven to be an effective method for probing various phases of the ISM at low and high redshift: the warm neutral ($T\la10^4$ K) component, using \ion{Na}{1} D [\citealt{phillips93} (NGC 1808); \citealt{heck00}]; the warm ionized ($T\sim10^{4-5}$ K) component, using UV lines [\citealt{lequeux95} (Mrk 33); \citealt{hl97} (NGC 1705); \citealt{sahu97} (NGC 1705); \citealt{kunth98,gd98,pet01}]; and the hot coronal ($T\sim10^{5-6}$ K) component, using \ion{O}{6} $\lambda\lambda1032,\;1038$ [\citealt{heck01} (NGC 1705)].  The existence of outflows can be inferred from the presence of absorption lines that are blueshifted with respect to the systemic velocity of the host galaxy.  We apply this technique to the warm neutral gas in a sample of 12 ULIGs using the \ion{Na}{1} D doublet and measure individual outflow properties such as absorbing column density and outflow velocity.  \citet{heck00} (hereafter HLSA) have used this feature to study a sample of 27 luminous infrared galaxies (LIGs; $L_{\mathrm{IR}} > 10^{11}L_{\sun}$) and 5 ULIGs with $z<0.12$, but generally at lower resolution ($55-170\;\kmsm$).  We analyze the line profiles in our targets by means of a $\chi^2$ minimization fitting algorithm that allows for non-Gaussian intensity profiles, multiple absorbing components, and a covering fraction that is less than one.  Assuming that all absorption components with velocities greater than $\sim70\;\kmsm$ are outflowing, we use a simple model of a mass-conserving free wind to compute the corresponding mass outflow rates.  We also compute star formation rates from infrared luminosities.
 
The ratio of the total mass outflow rate to the corresponding global star formation rate in each object is a measure of the ``reheating efficiency,'' or the efficiency with which star formation reheats the surrounding ISM and produces bubbles and outflows \citep{mar99}.  Previous measurements of the reheating efficiencies in nearby dwarf galaxies, edge-on spirals, and lower-luminosity infrared galaxies indicates that this ratio is of order unity (\citealt{mar99}; HLSA).  We can also track the dependence of the reheating efficiency on the mass of the galaxy, as parameterized by observed $R$- and $K^{\prime}$-band luminosities and the kinematics of the host galaxy.  We can estimate the fraction of absorbing gas that escapes the galaxy and enters the intergalactic medium by comparing outflow velocities with escape velocities based on measured kinematics.  From the reheating efficiency and escape fraction for each object, we are able to compute the ``ejection efficiency,'' which is the ratio of the outflow rate of material that escapes into the intergalactic medium to the corresponding global star formation rate.  This last quantity is most directly related to the impact of ULIGs on the IGM.

The structure of this paper is as follows.  Section 2 describes the sample, observations, and data reduction.  In \S 3, we present an analysis of the \ion{Na}{1} D lines; measure kinematics, optical depths, and covering fractions; and compute absorbing column densities.  We also present equivalent width measurements of prominent absorption lines.  In \S 4, we discuss and interpret the results; we compute mass outflow and star formation rates and compare the results to previous studies; and we discuss the ultimate fate of the outflowing gas.  In \S 5, we summarize and conclude.  For all calculations, we assume $H_0 = 75\; \kmsm\;\mathrm{Mpc^{-1}}$ and $q_0 = 0$.  All wavelengths are vacuum wavelengths unless otherwise noted and are taken from \citet{mort91} and/or the NIST Atomic Spectra Database\footnote{The NIST Atomic Spectra Database is available at \url{http://physics.nist.gov/cgi-bin/AtData/main\_asd}.}.

\section{SAMPLE, OBSERVATIONS, AND DATA REDUCTION}

Our targets come from the IRAS 1 Jy survey \citep{ks98}.  The results of this survey are a complete sample, down to a flux level of $f_{\nu}(60\;\mathrm{\micron}) = 1\;\mathrm{Jy}$, of 118 ULIGs with Galactic latitude $|b|>30^{\circ}$ and declination $\delta > -40^{\circ}$.  These objects have redshifts of $z=0.02-0.27$ and are the brightest sources with luminosities in the range $\log(L_{\mathrm{IR}}/L_{\sun})=12.00-12.84$.  Using low-dispersion optical spectra of resolution 8.3 \AA\ (\citealt{kvs98}; \citealt*{vks99a}), we visually selected 11 objects with most or all of the following characteristics: (1) \ion{Na}{1} D absorption is deep and possibly blueshifted from systemic; (2) \ion{He}{1} $\lambda 5876$ emission is weak or absent; and (3) the \ion{Na}{1} D lines are unaffected by the O$_2$ A- and B-bands near 7620 \AA\ and 6880 \AA, respectively.  The first selection criterion is not necessarily a robust one, since covering fractions less than one are typical in the sample (\S \ref{cf}) and it is sometimes difficult to determine velocity shifts from systemic using the low-dispersion spectra.  Various other selection criteria were operative, as well.  For instance, the bulk of our observing window included only $30\%$ of the entire sample, and F$04313-1649$ was chosen solely for its comparatively high redshift.  We also tried to choose objects whose energy output is dominated by star formation, rather than an AGN (i.e. \ion{H}{2}-region-like objects and LINERs; \citealt{lvg99}).  We note that the first selection criterion may create a bias in the sample towards objects with high optical depths and velocities relative to systemic; this possibility should be noted when considering the results.  However, the limited window of visibility during the observations and the other selection criteria involved means that this is probably not a strong bias.

Table \ref{sprop} lists the objects we selected and some of their relevant properties.  The optical spectral types are distributed as follows: 2 \ion{H}{2}-region-like objects; 7 LINERs; and 3 Seyfert 2s.  The object F$10190+1322$ has two nuclei that are well-separated by 6.2 kpc; we classify each nucleus separately based on our data, although the overall spectral class is \ion{H}{2}-region-like \citep{vks99a}.  Based on $R$- and $K^{\prime}$-band images \citep*{vks02}, all of the other targets are mononuclear.  The same images reveal that some objects have tidal tails, and that all targets show some sign of a recent merger, such as morphological asymmetry.

Table 1 lists a twelfth object, F$12540+5708$ (or Mrk 231).  Mrk 231 is an optically classified Seyfert 1, and it contains high-velocity, broad, and deep absorption lines that are reminiscent of broad absorption line quasar spectra.  Because of its distinct quasar-like properties, we treat Mrk 231 separately from the rest of the sample, and discuss it only in Appendix A.

We observed the targets at the W.M. Keck Observatory on Keck II using the Echellette Spectrograph and Imager (ESI).  In the echellette configuration, ESI is a high-efficiency, moderately-high resolution spectrograph that operates in low, cross-dispersed orders.  It has a constant $11.4\;\kmsm\;\mathrm{pixel^{-1}}$ dispersion; a wavelength coverage of $0.39-1.10\;\micron$; and a pixel scale of $0 \farcs 128-0 \farcs 179\;\mathrm{pixel}^{-1}$ along the slit (increasing from blue to red orders).  ESI operates with a MIT-LL $2048\times4096$ CCD.  We chose a slit width of 1\farcs0 and a north-south position angle ($\mathrm{PA} = 0^{\circ}$) for all observations.  From fits to sky lines in the seven reddest orders in our data, we estimate a constant spectral resolution of $R\sim4600$ for our observations, which corresponds to 65 \kms.  Table \ref{sprop} lists the UT date of observation and the exposure time for each object, as well as the mean seeing for each night.  All observations were made in photometric conditions.  The resulting signal-to-noise ratio of the data for most targets is $20-40$ per pixel (at maximum sensitivity), although for the brightest targets it is much higher.

We performed the initial data processing using the MAKEE\footnote{MAKEE is available at \url{http://www2.keck.hawaii.edu:3636/realpublic/inst/common/makeewww/index.html}.} data reduction package, written by T. Barlow for use with HIRES spectra and adapted to reduce ESI data.  MAKEE performs the following automatically: bias subtraction, flat-fielding, order-tracing, sky subtraction, object extraction, and wavelength calibration (including a heliocentric velocity correction and an air-to-vacuum wavelength conversion).  MAKEE also produces error spectra based on Poisson statistics, which we propagated consistently through the rest of the reduction procedure and used as input for the minimization algorithms in our spectral fitting routine (\S\ref{fitting}).  In most cases, CuAr, HgNe, and Xe lamps were used simultaneously for wavelength calibration; sky lines were later used to apply a constant shift to the solution (the resulting residuals are generally $\pm0.2$ pixel).  The entire spatial profile was extracted for each object; this is the most practical treatment due to the difficulty of maintaining a constant spatial aperture over the large range of $z$ in our sample, the fact that most of the objects are distant and not well-resolved, and the need to maximize the signal-to-noise ratio of the spectra.  Flux calibration was done in IRAF using the stars HD 37129 and HD 118246 and corresponding flux points spaced at intervals of 0.7 \AA\ \citep{gv88}, avoiding those portions of the stellar spectra with deep absorption lines.  A 200 \AA\ slice of the continuum around each \ion{Na}{1} D feature (450 \AA\ in the case of Mrk 231) was fit with a low-order spline, and the result was divided into the original spectrum to produce a spectrum normalized to unity at the continuum level.  (We used a similar normalization procedure for other absorption lines that were fit; see \S \ref{otheral} and Appendix \ref{append}.)

\section{RESULTS}

This section and the following one present the analysis and discussion of our observations.  We do not include Mrk 231 in this discussion, except for a brief treatment of its redshift in \S \ref{skin} and Table \ref{z} and a mention of its velocity dispersion in Table \ref{kin}.  We discuss Mrk 231 separately in Appendix A.  We also treat the two nuclei of F$10190+1322$ separately when appropriate.

\subsection{Line Fitting} \label{fitting}

In the case where the members of a doublet are decently resolved and unblended with each other or other lines, it is possible to compute column density and covering fraction for the absorbing species as a function of velocity (e.g. \citealt{bs97,hamann97,arav99a,arav99b}; see Appendix A for an example using \ion{Ca}{2} H \& K in Mrk 231).  However, due to the small velocity separation of the \ion{Na}{1} D lines ($\sim 300\;\kmsm$), the two profiles are generally too blended with each other for this procedure.  Instead, we make the curve-of-growth assumption of a Gaussian velocity distribution and fit multiple components to each line profile.  This technique is fairly robust; even if the overall profile is a sum of many Gaussian profiles, the fit of the overall profile will yield roughly accurate column densities and a representative velocity spread (assuming that there are no kinematically narrow but optically thick lines, or unresolved saturated structures; \citealt{jenkins86,ss91}).  We account for the possibility of covering fractions that are less than one by assuming a covering fraction for each fitted component that is independent of velocity.

Figure \ref{spectra} displays a $\sim80\;\mbox{\AA}$ portion of the spectrum around the \ion{Na}{1} D line for our eleven program galaxies, smoothed using a boxcar average of width 3 pixels (corresponding to 34 \kms).  Overlaid on these spectra are our fits to the line profiles.  Dashed lines indicate the individual components of each fit.  The profiles were fit using a $\chi^2$ minimization routine within the IRAF-based interactive spectral fitting software SPECFIT{\footnote{SPECFIT is available at \url{www.pha.jhu.edu/$\sim$gak/specfit.html}.} \citep{gak94}.  We added an extra routine to SPECFIT in order to fit absorption doublets with a partial covering fraction.  Each fit consisted of several components: a flat continuum; one or more \ion{Na}{1} D absorption components (where one component equals one doublet pair); and, when necessary, a \ion{He}{1} $\lambda5876$ emission component.  Table \ref{mprop} lists the parameters that we measured for each absorption component.

The wavelengths in vacuum of the two lines of the \ion{Na}{1} D doublet are $\lambda_1=5897.558$ \AA\ and $\lambda_2=5891.583$ \AA\ \citep{mort91}, and their oscillator strengths are (to $1\%$ accuracy) in the ratio 2:1, respectively, so that the optical depth at the centroid of the blue line is twice that at the centroid of the red line (i.e. $\tau_{2,c} = 2 \tau_{1,c}$).  For each doublet component, we fit four parameters: the central optical depth of the weaker, red line, $\tau_{1,c}$; the centroid of the weaker line, $\lambda_1$; the full width at half-maximum  (FWHM) of the physical velocity distribution of the absorbing gas; and a covering fraction, $C_f$.  The central optical depth of the stronger, blue line, $\tau_{2,c}$, then follows from the above relation, and $\lambda_2$ follows from the ratio of the two rest wavelengths, which is independent of redshift and relative velocity.  

Assuming a continuum level of unity and a covering fraction $C_f$, the intensity (measured from zero) across the doublet as a function of wavelength is
\begin{equation}
I(\lambda) = 1 - C_f(1 - \exp\{-[\tau_1(\lambda)+\tau_2(\lambda)]\}),
\end{equation}
where $\tau_1(\lambda)$ and $\tau_2(\lambda)$ are the optical depths in the weak and strong lines, respectively, as a function of wavelength.  The curve-of-growth assumption of a Gaussian in velocity space leads to a Gaussian in optical depth for each doublet line.  Substituting this optical depth profile into the above equation and using the relationship between $\tau_{1,c}$ and $\tau_{2,c}$ leads to
\begin{equation}
\label{intensity}
I(\lambda) = 1 - C_f(1 - \exp\{-\tau_{1,c}[e^{-(\lambda-\lambda_1)^2/(\lambda_1 b/c)^2}+2e^{-(\lambda-\lambda_2)^2/(\lambda_2 b/c)^2}]\}),
\end{equation}
where $b$ is the Doppler parameter in \kms\ [$b=\sqrt{2}\sigma=\mathrm{FWHM}/(2\sqrt{\ln 2})$] and $c$ is the speed of light.  This equation applies in both the rest frame and the observer's frame.

Note again that we make the assumption that the covering fraction $C_f$ is independent of velocity, and hence $\lambda$, for a given component.  $C_f$ is also technically an ``effective'' covering fraction in the sense that it describes the fractional area of the background light source that is obscured by the absorbing gas {\it as well as} any continuum light that may be scattered into the line of sight.  Two or more absorbing components within a given object may have different covering fractions; for simplicity in the fitting, we assumed that for $n$ components, each having intensity $I_i$ as given by equation (\ref{intensity}), the total intensity reaching the observer is $I_{\mathrm{tot}} = \prod_{i=1}^n\;I_i$.  Modulo the scattered light, this means physically that $C_f$ for a given component describes not only the covering fraction of the background light, but also the fraction of other components that it overlaps with (although the material can be either behind or in front of these other components).  The gas can be distributed arbitrarily along the line of sight.

The \ion{He}{1} $\lambda 5876$ emission line was fit with a Gaussian intensity profile.  In five objects there was no obvious emission, so no line was fit.  This component has two free parameters: the FWHM of the Gaussian and the integrated flux under the line.  In all cases, we constrained the line center to be at or near the systemic velocity of the galaxy.  Again, for simplicity in the fitting, we assumed that $I_{\mathrm{tot}} = I_{\mathrm{emission}} \times I_{\mathrm{absorption}}$, where $I_{\mathrm{emission}}$ and $I_{\mathrm{absorption}}$ are both measured from zero; this means that $C_f$ also describes the fractional coverage of the line-emitting region.

There may be another important consideration in the line fitting.  The \ion{Na}{1} D line in the spectra of galaxies is produced by both stellar and interstellar absorption, but we are interested solely in the interstellar component. To estimate the magnitude of the stellar contribution to the line in each object, we measured the equivalent width of the \ion{Mg}{1} b $\lambda\lambda 5167,\;5173,\;5185$ feature (Table \ref{alew}), which is mostly stellar in origin.  Furthermore, Mg and Na are produced in roughly the same nuclear processes in hot stars, suggesting that there may be a correlation in the abundance of these two elements within stellar atmospheres.  To determine this correlation, we used equivalent width data from \citet*{hbc80} on the nuclei of nearby bright galaxies and from \citet{ba86} on Galactic globular clusters and a sample of mostly non-active galaxies.  Using these data sets, we determined that $W_{\mathrm{eq}}^{\star}$(\ion{Na}{1} D) $\sim\;0.5$ $W_{\mathrm{eq}}^{\star}$(\ion{Mg}{1} b), with a possible intrinsic scatter of $\ga\;25\%$.  The resulting estimates of the fractional contribution of stellar absorption to the \ion{Na}{1} D lines in these objects are small (see Table \ref{mprop}); they are in the range $4-28\%$, and have a median and mean value of $15\%$ [with a $\sim2$-$3\;\sigma$ error of $\pm5\%$ due to uncertainty in the measurement of $W_{\mathrm{eq}}^{\star}$(\ion{Mg}{1} b)].  Due to the difficulty of modelling high-resolution profiles of stellar absorption lines that are integrated across an entire galaxy and the fact that the stellar contribution is quite small, we chose to fit the \ion{Na}{1} D lines without subtracting the stellar absorption or including a stellar component in our fits.  The stellar contribution to the line will likely increase the measured optical depths and covering fractions but decrease the measured blueshift; these effects will, to some degree, offset each other in the estimation of mass outflow rates (eq. [\ref{mdotf}]).

\subsection{Kinematics} \label{skin}

The heliocentric redshifts listed in Tables \ref{sprop} and \ref{z} were determined from the current optical dataset using prominent emission and absorption lines, except in the cases of F$05189-2524$ and Mrk 231, for which we used the more precise redshifts determined from CO emission \citep*{sss91} and \ion{H}{1} absorption \citep*{cwu98}, respectively.  We measured separate redshifts for stellar absorption lines and nebular emission lines and averaged the two if they differed by no more than $\Delta z = \pm0.0002$; otherwise, we adopted the absorption-line redshift.  Possible errors include: the 65 \kms\ resolution of our spectra; the lower signal-to-noise ratio in those parts of the spectrum where the absorption lines fall; and systematic uncertainties due to unusual kinematics in a few cases.  After accounting for these errors, the optical redshifts have a $\sim2$-$3\;\sigma$ uncertainty of $\sim0.0002$ (or 60 \kms).  The vertical dotted lines in Figure \ref{spectra} indicate the wavelengths of \ion{Na}{1} D and \ion{He}{1} $\lambda5876$ at the systemic velocity of each galaxy determined from these redshifts.

In Figure \ref{dvhist}, we plot the distributions of five kinematic quantities measured from the \ion{Na}{1} D profile fits.  The velocity centroids relative to systemic for each component are
\begin{equation}
\Delta v \equiv v_{\mathrm{sys}} - v,
\end{equation}
where $v_{\mathrm{sys}}$ is the heliocentric velocity of the host galaxy and $v$ is the heliocentric velocity of the centroid of the absorbing gas component.  (Note that we use the opposite of the usual convention---velocities blueshifted relative to the host galaxy are positive.)  A ``maximum'' blueshifted velocity for each component is given by
\begin{equation}
\Delta V = \Delta v + \onehalf \mathrm{FWHM}.
\end{equation}
An interesting subset of these two distributions is the maximum $\Delta v$ and $\Delta V$ for each object, given by
\begin{eqnarray}
\Delta v_{\mathrm{max}} &=& max(\Delta v)\\
\mathrm{and} \; \Delta V_{\mathrm{max}} &=& max(\Delta V).
\end{eqnarray}
(``Maximum'' here accounts for sign---in other words, it describes the most blueshifted velocity, or least redshifted velocity if no blueshifted velocity exists.)  A final kinematic quantity is the distribution of Doppler parameters $b$ for all components.  The measured values for FWHM are corrected in quadrature for the instrumental resolution of 65 \kms\ before conversion to $b$ (the same correction is made throughout this paper for FWHM).  Values for $\Delta v$, $\Delta V$, and $b$ for each component are also listed in Tables \ref{mprop} and \ref{cprop}.  The $\sim2$-$3\;\sigma$ uncertainty in these measurements varies from source to source, but is on the order of $40-60\;\kmsm$(due to uncertainties in the measured redshifts, the effects of finite instrumental resolution, and uncertainties in the fitting).

The median and mean values, respectively, of each of these distributions are as follows: (a) $\Delta v$, 104 \kms\ and 233 \kms; (b) $\Delta V$, 268 \kms\ and 444 \kms; (c) $\Delta v_{\mathrm{max}}$, 284 \kms\ and 381 \kms; (d) $\Delta V_{\mathrm{max}}$, 511 \kms\ and 636 \kms; and (e) $b$, 200 \kms\ and 254 \kms. Most components (12 of 19, or $63\%$) have velocities $\Delta v$ that are blueshifted more than $70\;\kmsm$ from systemic---i.e. $\Delta v > 70\;\kmsm$---and most objects (8 of 11, or $73\%$) have a maximum velocity $\Delta v_{\mathrm{max}} > 70\;\kmsm$.  This percentage can be compared to other recent results.  In their study of 32 local LIGs, HLSA find that $38\%$ of their sample hosts outflows, while \citet{pet01} find evidence for outflows in $100\%$ of their 19-member, $z \sim 3$ sample.

For the purposes of calculating mass outflow rates (\S \ref{mor}), we assume that most blueshifted components represent outflowing material.  There are two blueshifted components within $15\;\kmsm$ of systemic; we assume this is quiescent interstellar material.  However, the other blueshifted components have velocities in excess of $70\;\kmsm$.  Note that stellar contamination of the line will bias toward $\Delta v = 0\;\kmsm$ those near-systemic components that are truly outflowing.  Note also that substantial systematic errors in redshift could lead to larger than expected errors in $\Delta v$.  Another important consideration is the fact that most or all ULIGs are mergers in late stages of interaction \citep{sm96,vks02}.  The interstellar medium in some of these galaxies may then contain tidal material leftover from earlier stages, which could account for the redshifted material observed in a few objects (material with $\Delta v < -100\;\kmsm$ is observed in two objects).  This may also account for a small portion of the blueshifted material that we attribute to outflows.  Unfortunately, it is not possible to distinguish these two effects without spatial information.

The Doppler parameter $b$ describes the kinematic spread of the absorbing material.  Assuming that the absorber is gas that is entrained in a wind, the rather large values that we measure for $b$ are likely due to the fact that we are simultaneously detecting material from all parts of the outflow.  In other words, unresolved projection effects will produce a large range of radial velocities along our line of sight, and multiple components with smaller values of $b$ may blend together and mimic a single component with a large $b$ \citep{jenkins86}.  The case of a fragmented superbubble (in the blowout stage of a galactic wind) or a clumpy ISM would be consistent with this.  Turbulent motion created by shocks in the outflow will also contribute to the observed velocity widths.

In \S \ref{mor} we assume that the outflowing gas motions are purely radial.  It is also likely, based on observations of local galactic winds \citep[e.g.][]{ham90,veil94,sbh98,vr02}, that the solid angle into which the wind is outflowing is less than $4 \pi$, typically due to a collimated, biconical wind structure.  If we then assume a point source for the background continuum, our velocity measurements represent actual outflow velocities.  However, if the background source is extended and we are not looking down the axis of the assumed biconical wind structure, then projection effects may cause our measurements to underestimate the actual outflow velocities.

\subsection{Optical Depth and Column Density} \label{columndensities}

The central optical depth of the weak line, $\tau_{1,c}$, for each component is listed in Table \ref{mprop}, and the distributions of this quantity for all components and only those components assumed to be outflowing are plotted in Figure \ref{cfhist}$a$.  The median and mean of these distributions are 0.42 and 0.83, and 0.39 and 0.97, respectively, and the $\sim2$-$3\;\sigma$ uncertainty in $\tau_{1,c}$ is $\sim25\%$.  The corresponding \ion{Na}{1} column densities for each component are listed in Table \ref{cprop} and plotted in Figure \ref{nhist}.  We also show in Figure \ref{nhist} the column densities for only outflowing components and the total column density for each object, summed over only outflowing components.  These column densities are computed from $\tau_{1,c}$ and $b$ using \citep{spitz78}
\begin{equation}
N(\mbox{\ion{Na}{1}})=\frac{\tau_{1,c} b}{1.497\times10^{-15} \lambda_1 f_1},
\end{equation}
where $f_1=0.3180$ is the oscillator strength of the weak line \citep{mort91}, $b$ is in \kms, and $\lambda_1$ is in \AA\ (in the rest frame, in vacuum).  The median and mean values for $N$(\ion{Na}{1}) are: (a) for each component, $3.0\times10^{13}\;\mathrm{cm}^{-2}$ and $4.3\times10^{13}\;\mathrm{cm}^{-2}$; (b) for only components assumed to be outflowing, $3.3\times10^{13}\;\mathrm{cm}^{-2}$ and $5.0\times10^{13}\;\mathrm{cm}^{-2}$; and (c) for each object, $3.6\times10^{13}\;\mathrm{cm}^{-2}$ and $5.5\times10^{13}\;\mathrm{cm}^{-2}$.

In Table \ref{mprop}, we also list the doublet equivalent width ratio for each absorbing component, $R = W_\mathrm{eq}^{\lambda5890} / W_\mathrm{eq}^{\lambda5896}$.  These values are not actually measured, but rather computed using synthetic profiles based on the measured profile parameters.

Note that if the wind is collimated and the axis of the outflow is nearly in the plane of the sky, it is possible that very little of the wind will be in front of a strong background continuum source.  This could be true for some of the galaxies in our sample, in which case the real distributions of $\tau$ and $N$(\ion{Na}{1}) for outflows in these objects will be higher than we measure.

On average, our measured optical depths are a factor of $\sim4-10$ less than those measured by HLSA in the \ion{Na}{1} D line for their sample of 32 starburst-dominated LIGs and ULIGs.  For their subsample in which the \ion{Na}{1} D doublet is dominated by interstellar absorption, HLSA find a median value of $\tau_{1,c} \sim 4$ using the doublet ratio method and $C_f = 1$.  For three objects with cleanly resolved doublets and $C_f \sim 0.8$, they find $\tau_{1,c} \sim 2$.  However, the values that they infer for  $N$(\ion{Na}{1}) are in the range $(5-10)\times10^{13}\;\mathrm{cm^{-2}}$, which is comparable to our column density distribution.

The details of our fitting algorithm relative to that of HLSA may be the cause of the discrepancies among our values for $\tau_{1,c}$.  We use a different program (``specfit'' vs. ``splot''), fit one to three components, and fit Gaussians in optical depth, rather than intensity.  An alternative explanation for the differences in the two datasets is a real physical difference in the two samples, which is a possibility given that ours is composed only of ULIGs, while the HLSA sample consists mainly of lower-luminosity galaxies.  It should again be emphasized, however, that the resulting total column densities are similar in each case.

Two methods exist for converting $N$(\ion{Na}{1}) to a hydrogen column density.  The first (``method 1'') requires making the proper ionization and depletion corrections separately.  For cool gas along the line-of-sight to $\zeta$ Oph ($l = 6 \fdg 28,\;b = +23 \fdg 59,\;d = 140\;\mathrm{pc})$, \citet{mort74,mort75} uses a simple photoionization model with electron density $n_e=0.7 \; \mathrm{cm}^{-2}$ and kinetic temperature $T=56 \; \mathrm{K}$ to compute $N$(\ion{Na}{2})/$N$(\ion{Na}{1}) $=3.1$.  The corresponding depletion value, taken in reference to the total neutral atomic plus molecular gas [$N$(\ion{H}{1})$ + 2 N$(H$_2$)], is then $-0.90$.  [\citet{ss96} use different solar reference abundances and infer a depletion value of $-0.95$.]  Data from \citet*{ppg84} along sightlines toward nine stars near the Galactic plane indicate that the depletion correction for Na is independent of the line-of-sight gas density, suggesting that this value is fairly robust.  However, the ionization correction will be more dependent on local physical conditions, and is almost certainly larger than 3.1 in our objects (especially considering that the temperature is probably at least two orders of magnitude larger than 56 K).  In general, the physical conditions in outflowing gas in ULIGs are likely to be harsher than along sightlines through the local Galactic ISM.  Specifically, shock ionization due to the fast motion of the wind through the quiescent ambient medium and photoionization due to a background continuum from stars, an AGN, and/or hot wind material are likely to produce a higher ionized fraction of Na (and H).

A second method (``method 2'') uses an empirically-determined factor to convert directly from $N$(\ion{Na}{1}) to $N$(H)---also derived using data from sightlines through the local Galactic ISM---thus performing the depletion and ionization corrections at once.  The relation is of the form
\begin{equation}
\log[N(\mathrm{Na\;\mbox{\small I}})] = \beta \log[N(\mathrm{H})] + \gamma,
\end{equation}
where N(H) may refer to only neutral atomic gas (\ion{H}{1}) or both atomic and molecular gas, and column densities are in units of cm$^{-2}$.  In Table \ref{natoh}, we compare values for $\beta$ and $\gamma$ from the literature and the size of the conversion factor relative to method 1 [i.e. the ratio of $N$(H) determined using each method].  In all cases where both $\beta$ and $\gamma$ are available, method 2 produces a larger correction by factors ranging from 5 to 41.  The correction factor is not uniform across our dataset, either, except for values of $\beta$ close or equal to unity.  In other words, for values of $\beta$ not equal to unity, the correction factor probably has a dependence on density, $n_{\mathrm{H}}$.  (The specific value of $\beta$ may indicate something about the ionization of the ISM; see \citealt{hobbs74a,hobbs76}.)  For our work, we choose the conversion relation derived by \citet{stokes78}.  It is similar to the earlier results of \citet{hobbs74a,hobbs74b,hobbs76}, and is partly based on this work.  The value derived by \citet*{fvg85} for $\beta$ is probably too small, largely due to the subdivision of $N$(H) along some lines of sight into multiple kinematic components \citep{herbig93}.  \citet{herbig93} includes no values for $\gamma$, unfortunately, making his values for $\beta$ impossible to use.

Our choice of method 2 with $\beta = 2.11$ and $\gamma = -31.3$ means that our values for $N$(H) are a factor of $6-29$ higher than those determined from the depletion correction alone [with the actual value dependent on $N$(\ion{Na}{1})], or, if we include an ionization correction of 3.1 in method 1, a factor of $2-9$ higher.  In other words, the effective ionization correction that we make using method 2 is much larger than in method 1.  Our values for $N$(H) are at most a factor of two larger than those of HLSA (although the uncertainty of this conversion applies to their results, as well).  However, we still do not know how physical conditions in the gas from which this correction is determined compare to those in the absorbing gas in galactic winds.  As discussed above, the conditions in galactic winds are likely to produce a higher level of ionization, suggesting that these estimates for $N$(H) are still a lower limit to the real values.

\subsection{Covering Fraction} \label{cf}

The measured covering fraction for each component is listed in Table \ref{mprop} and the distributions for all components and only those components assumed to be outflowing are plotted in Figure \ref{cfhist}$b$.  The median and mean of these distributions are 0.61 and 0.59, and 0.61 and 0.56, respectively.  The measurement error in this quantity is $\sim0.1$.

As discussed in \S\ref{fitting}, the covering fraction gives information about the spatial distribution of the gas, as well as any light that is scattered into the line of sight.  Even in the case that it describes solely the amount of absorbing material in front of the continuum source, there are still degenerate configurations which are impossible to distinguish without more spatial information.  If the size of the background light source is small compared to the global size of the absorbing gas structure, then the covering fraction is likely to describe the local clumpiness of the absorbing material.  However, if the background light is much more extended, then it will also contain information about the global angular size of the absorbing region with respect to the background light.

We state in \S \ref{columndensities} that if the outflow is collimated and its axis lies nearly in the plane of the sky, then the outflow may cover very little of the galactic continuum source.  Lower covering fractions could also result from this inclination effect.

\subsection{Other Blueshifted Absorption Lines} \label{otheral}

There are other absorption lines of interest which potentially contain information about the outflows in these objects, such as \ion{Ca}{2} H \& K and \ion{K}{1} $\lambda\lambda7665,\;7699$.  The \ion{K}{1} doublet typically has a very low signal-to-noise ratio because \ion{K}{1} is much less abundant than \ion{Na}{1} and the feature is typically redshifted into areas of our spectra where sensitivity is low and night sky lines are prominent.  However, the stronger doublet line has a high enough signal-to-noise ratio in six objects to make a measurement of the velocity centroid and gaussian FWHM possible.  For $\sim3$ of these objects (F$03250+1606$, F$09116+0334$, and F$11387+4116$), both the offsets and widths match those determined for one \ion{Na}{1} component to within $\sim30\%$.  However, there are larger discrepancies for the other objects; we attribute these errors to the existence of multiple components and the low signal-to-noise ratio of the \ion{K}{1} lines.  Because the weak line of the doublet has too low a signal-to-noise ratio to be useful, we are unable to use it as a diagnostic in any object and cannot constrain the \ion{K}{1} optical depth or covering fraction.

The \ion{Ca}{2} H \& K lines are highly contaminated by stellar absorption for low values of $\Delta v$.  The datasets that we used to infer the relationship between the equivalent widths of \ion{Mg}{1} b and \ion{Na}{1} D \citep{hbc80,ba86} imply that $W_{\mathrm{eq}}^{\star}$(\ion{Ca}{2} K) $\simeq$ 2 $W_{\mathrm{eq}}^{\star}$(\ion{Mg}{1} b).  Based on our measured absorption line equivalent widths (Table \ref{alew}), we then expect the average stellar contribution to \ion{Ca}{2} K to be $\sim35-40\%$; since this is a substantial fraction of the line, we are unable to treat it as primarily interstellar in origin.  However, in one instance, a high-velocity component is well-separated from the near-systemic components, the $\Delta v = 1540\;\kmsm$ line in F$05024-1941$.  Although this part of the spectrum has a lower-than-average S/N, we were able to fit the profiles of these lines (fixing the covering fraction to be the same as for \ion{Na}{1} D).  The results are as follows: $\tau_{1,c} = 0.6$; $\lambda_1 = 4713.1\;\mbox{\AA}$; $b = 104\;\kmsm$; $\Delta v = 1559\;\kmsm$; and $N$(\ion{Ca}{2})$=2.9\times10^{13}\;\mathrm{cm}^{-2}$.  Note that the values of $\Delta v$ determined from \ion{Ca}{2} and \ion{Na}{1} agree to within the specified errors.  However, extrapolating to a hydrogen column density is difficult, largely because of the strong dependence of Ca depletion on density \citep{ppg84}, which may contribute to the fact that $N$(\ion{Ca}{2}) and $N$(H) are not well-correlated along local Galactic sightlines \citep{hobbs74b}.  The depletion for Ca measured toward $\zeta$ Oph is $-3.73$ \citep{ss96}, a full factor of $10^3$ larger than that measured for Na ($-0.95$).  Using this value yields a larger $N$(H) than that measured from Na, but it is likely that Ca is actually less depleted than this.  For instance, if dust grains are destroyed by the shocks present in these outflows \citep{ppg84,ss96}, then the Ca depletion and $N$(H) determined from Ca will both decrease.

\subsection{Absorption Line Equivalent Widths} \label{eqw}

We measured the equivalent widths of the absorption lines used in this paper using ``splot'' in IRAF; the results are listed in Tables \ref{mprop} and \ref{alew}.  We quote rest-frame equivalent widths, which are related to the observed equivalent widths by $W_{\mathrm{eq}}^{\mathrm{rest}} = W_{\mathrm{eq}}^{\mathrm{obs}}/(1+z)$.  To measure the \ion{Na}{1} D equivalent widths, we used the same continuum-normalized sections as in the line fitting.  We used the command in ``splot'' that directly measures $W_{\mathrm{eq}}$ given two endpoints, selecting the boundaries by eye (and stopping where the line rose to the continuum level or the \ion{He}{1} emission line).  For the other lines, the measurements were generally made using a Gaussian fit in ``splot,'' with no prior continuum fitting.  We made no corrections due to broadening of the lines by the intrinsic velocity dispersions of the objects \citep*[see][]{tdt90}.  The \ion{Na}{1} equivalent widths are reasonably accurate ($\pm5\%$), while those for other lines have a higher uncertainty, roughly $\pm20\%$ due to uncertainties in continuum placement, lack of sensitivity in the blue and red portions of the spectrum where the lines were typically found, and the presence of sky lines (in the case of the \ion{Ca}{2} absorption triplet, $\lambda\lambda 8498,\;8542,\;8662$).  The strengths of the observed lines (the \ion{Ca}{2} triplet, \ion{Ca}{2} K, the Balmer lines, and \ion{Mg}{1} b) suggest that they are produced by a young stellar population with red supergiants superimposed on an older population (\citealt{tdt90}; \citealt*{bas90}).

\section{DISCUSSION} \label{discussion}

\subsection{Mass Outflow Rate} \label{mor}

The primary purpose of this paper is to estimate total mass outflow rates from the ULIGs in our sample and compare them to the corresponding global star formation rates.  We compute mass outflow rates from our measurements and a simple model.  Consider a mass-conserving free wind (i.e. the mass outflow rate $\dot{M}$ and velocity $v$ are independent of radius $r$) in a medium of number density $n(r)$.  Assume that the wind flows from an inner radius $r_{\star}$ to infinity, where $r_{\star}$ may correspond to the size of the wind-producing region.  The mass outflow rate at a radius $r$ is given by
\begin{equation} \label{m(r)}
\dot{M} = \Omega C_\Omega r^2 n(r) v m,
\end{equation}
where $\Omega$ is the solid angle subtended by the wind, $C_\Omega$ is the angular covering fraction of the gas within the solid angle $\Omega$, and $m$ is the mean particle mass.  The total observed column density in the outflow is the integral of $n$ along our line-of-sight:
\begin{equation} \label{Nint}
N = \int_{r_{\star}}^{\infty} n(l) dl.
\end{equation}
Solving equation (\ref{m(r)}) for $n(r)$, substituting this into equation (\ref{Nint}), and integrating leaves us with
\begin{equation}
\dot{M} = \Omega C_\Omega r_{\star} N v m.
\end{equation}
Normalizing to values typical to our sample and using $m = m_p$ and $C_\Omega = C_f$, we have 
\begin{equation} \label{mdotf}
\dot{M}(\mathrm{H}) = 21\;\left(\frac{\Omega}{4\pi}\right)\;C_f\;\left(\frac{r_{\star}}{1\;\mathrm{kpc}}\right)\;\left(\frac{N(\mathrm{H})}{10^{21}\;\mathrm{cm}^{-2}}\right)\;\left(\frac{\Delta v}{200\;\kmsm}\right)\;\smpym.
\end{equation}
\citet{ms89} find that the ratio of molecular to atomic gas in luminous and ultraluminous infrared galaxies increases with increasing far-infrared excess (the ratio of infrared to blue flux), and that ULIGs are likely to have more molecular than atomic gas by a factor of at least $1-10$.  It is unclear how much molecular gas is entrained in galactic outflows, however, so we choose $m = m_p$ as a conservative estimate, rather than $m = 2 m_p$.  Note also that $C_f$ is distinct from $C_\Omega$ in that $C_f$ describes the line-of-sight (rather than angular) covering fraction, and may include information about $\Omega$.  However, we use it as a reasonable estimate for, and probably a lower limit to, $C_\Omega$, especially given the probability of inclination effects in our sample.

We compute $\dot{M}$ for each absorbing component with $\Delta v > 70\;\kmsm$ using $\Omega = 4\pi$ and $r_{\star} = 1\;\mathrm{kpc}$ and list the results in Table \ref{cprop}; we assign $\dot{M} = 0\;\smpym$ to all other components.  We also list in this table the total mass outflow rate for each object, $\dot{M}_{\mathrm{tot}}$.  The distribution of $\dot{M}$ for each component is plotted in Figure \ref{mhist}$a$, with a median and mean of  $13\;\smpym$ and  $24\;\smpym$; the distribution of $\dot{M}_{\mathrm{tot}}$ is shown in Figure \ref{mhist}$b$, with a median and mean of  $22\;\smpym$ and  $41\;\smpym$.

The dominant source of error in $\dot{M}$, besides our choice of model and any unresolved spatial details, is probably the uncertain conversion of $N$(\ion{Na}{1}) to $N$(H) (\S \ref{columndensities}).  However, as discussed above, we likely underestimate the actual values of $N$(H) and $m$, suggesting that $\dot{M}$ may be a lower limit.  Other sources of error include our inability to measure $r_{\star}$ and $\Omega$.  Note that $\Omega < 4\pi$ for many local examples of galactic winds; a change in this quantity will reduce our estimate of $\dot{M}$ by at most a factor of a few.  However, we may also underestimate $C_\Omega$ by using $C_\Omega = C_f$, and $C_f$ may include a contribution from $\Omega$ due to inclination effects (see above and \S \ref{cf}).  This will mitigate any errors due to our use of $\Omega = 4\pi$.  Furthermore, $N$(\ion{Na}{1}) could also be an underestimate due to these inclination effects (\S \ref{columndensities}).

\subsection{Star Formation Rate and Reheating Efficiency} \label{sfr}

Table \ref{cprop} also lists the star formation rate calculated for each of the galaxies in our sample.  The star formation rate is proportional to $L_{\mathrm{IR}}$ according to \citep{kenn98}
\begin{equation}
\mathrm{SFR}= \alpha\;\frac{L_{\mathrm{IR}}}{5.8\times10^{9}\;L_{\sun}}.
\end{equation}
This equation is based on the stellar synthesis models of \citet{lh95} and assumes the following: a continuous burst of star formation of length $10-100$ Myr (the mean bolometric luminosity over this age range is used); solar abundances; a Salpeter IMF with mass cutoffs of 0.1 and 100 $M_\sun$; and that the dust reradiates the bolometric luminosity of the starburst in the infrared (i.e. $L_{\mathrm{IR}} \simeq L_{\mathrm{bolometric}}$).  We have included a correction factor $\alpha$ that accounts for the fraction of the bolometric luminosity of the galaxy that is produced by star formation (i.e. $0 \leq \alpha \leq 1$).

Mid- and far-infrared ISO spectroscopy appears to indicate that, on average, the bolometric luminosity of ULIGs is dominated by star formation, rather than an AGN \citep{genzel98}.  A two-dimensional diagnostic diagram composed of the strength of the 7.7 \micron\ PAH feature and mid-infrared line ratios suggests that $70-95\%$ of the luminosity of a typical ULIG results from star formation, while the rest comes from an AGN \citep{genzel98}.  This diagram can be used to roughly estimate the fraction of star formation and AGN activity that contribute to $L_{\mathrm{IR}}$.  However, the exact numbers are quite uncertain, and thus should be applied with care.  Since ISO spectroscopy exists for only one of the objects in our sample (F$05189-2524$), we use the available optical spectroscopic classifications (Table \ref{sprop}), which indicate \ion{H}{2}-region-like or LINER classifications for most of our objects, to infer that $\alpha\sim0.8$ is generally applicable to our sample.  [Spectral classifications based on optical emission-line ratios and the mid-infrared diagram (or simply the 7.7 \micron\ strength) match remarkably well, and comparison of the two suggests that LINERs are starburst-dominated objects \citep{lvg99}.]  In objects that are optically classified as Seyfert 1s or 2s, the downward correction may be larger, perhaps $\alpha\sim0.2-0.6$. For F$05189-2524$, we choose $\alpha = 0.4$, and for the other two Seyfert 2s, $\alpha = 0.6$.  We apply $\alpha = 0.4$ to F$05189-2524$ (an optical Seyfert 2) because it has an obscured broad-line region and has been shown to host a dominant AGN \citep{vks99b}.  Furthermore, \citet{laur00} give the strength of its 7.7 \micron\ feature as 0.4, which is identical to that measured by \citet{genzel98} for Mrk 231, an optically-classified Seyfert 1.

The total mass outflow rate in each object normalized to the corresponding global star formation rate,
\begin{equation}
\eta \equiv \frac{\dot{M}_{\mathrm{tot}}}{\mathrm{SFR}},
\end{equation}
is labeled the ``reheating efficiency'' by \citet{mar99}.  We list $\eta$ for each object in Table \ref{cprop}.  For the 8 galaxies hosting a wind, $\eta$ is in the range $0.07-0.66$, with a median and mean of 0.24 and 0.30.  For all 11 objects, the median and mean of $\eta$ are 0.14 and 0.22.  These values are lower on average by a factor of $\sim10$ than those measured by \citet{mar99} for warm ionized and hot X-ray-emitting gas in a sample of ten dwarf and six edge-on spiral galaxies, which have $\eta \sim 0.7-5$.  HLSA find that the reheating efficiencies for their sample are of order unity, although they do not measure specific values of $\dot{M}$ for each object.

The global star formation rates in the dwarfs and edge-on spirals surveyed by \citet{mar99} are much smaller than those in ULIGs, however.   The average star formation rate for the dwarf galaxies in the \citet{mar99} sample is $\sim0.08\;\smpym$, based on integrated H$\alpha$ fluxes \citep{mar98}; this is $3-4$ orders of magnitude smaller than the typical star formation rate in ULIGs.  The six edge-on galaxies also have lower star formation rates by about two orders of magnitude, based on infrared luminosities from \citet{rand96} (although they have areal star formation rates that are comparable to those of dwarf galaxies, when averaged over the entire galaxy; \citealt {mar99}).  The galaxies in HLSA that contain outflows (including three ULIGs) have a median infrared luminosity of $L_{\mathrm{IR}}=2\times10^{11}\;L_{\sun}$, which corresponds to a global SFR of $34\;\smpym$, an order of magnitude smaller than in our sample.  Finally, the global averaged surface brightnesses in ULIGs (over sizes of a few hundred parsecs, derived from mid-infrared observations) are typically much larger, by factors $\sim100$, than those measured for typical starburst galaxies, including those mentioned above \citep{soifer00,meurer97}.

The dense concentrations of molecular gas in ULIGs are likely to provide a large amount of dynamical resistance to any outflow that must travel through them.  ULIGs typically contain large amounts of molecular gas in their nuclei, on the order of $10^{10}\;M_\sun$, on scales of less than 1 kpc \citep[e.g.][]{ds98}.  The gravitational potential of these dense nuclei may restrict the outflow, as well.

In light of these considerations, it is remarkable that the reheating efficiencies in these ULIGs are only a factor of 10 lower than those in less massive galaxies.  At face value, our results imply that the outflow rate of a galactic wind scales (to within an order of magnitude) with the star formation rate of its host galaxy over a wide range of SFR.  Some of the difference in the average $\eta$ between our study and that of \citet{mar99} may be consistent with our suggestion that our values of $\dot{M}_{\mathrm{tot}}$ are really lower limits to the actual values.  However, it is also possible that the difference in $\eta$ between the two samples results from the different physical conditions in ULIGs.  Finally, we should note that the interstellar gas phases traced by the study of \citet{mar99} are hotter and more ionized than the ``cool'' ($T\sim10^{3-4}\;K$) neutral gas to which our survey is sensitive.  Thus, the discrepancy may also arise from having more outflowing gas in hotter, ionized gas phases.  The relative amount of material in the different interstellar gas phases of an outflow is still uncertain even in nearby objects; future multi-phase studies should shed light on this question.

\subsection{Host Galaxy Kinematics, Escape Fraction, and Ejection Efficiency} \label{hg}

The fraction of outflowing gas that escapes into the intergalactic or intracluster medium is an important parameter in outflow models for the mass-metallicity relation, the enrichment of the IGM, and galaxy formation.  This escape fraction could be measured by comparing estimates of the escape velocity for each galaxy to the outflow velocity of each component.

\subsubsection{Host galaxy kinematics}

We parameterize the mass and kinematics of each galaxy using five methods: (1) absolute $R-$ and $K^{\prime}$-band magnitudes (\citealt{vks02}; see Table \ref{sprop}), which trace the stellar populations of each galaxy; (2) the central line-of-sight velocity dispersion, determined from the fundamental plane of ellipticals and often used to represent random stellar motions; (3) the average emission line FWHM and the average FWHM of the \ion{Ca}{2} triplet ($\lambda\lambda8498,\;8542,\;8662$), probing gaseous and stellar kinematics, respectively, and representing a superposition of random and bulk circular motions; (4) position-velocity diagrams that indicate rotation of the emission-line gas, which are available in a few cases from our spectra; and (5) rotation curves from integral-field spectroscopy of the Pa$\alpha$ line that exist for the two nuclei in F$10190+1322$ \citep{murphy01}.  See Table \ref{kin} for a summary of host galaxy kinematics.

In method 1, the $R$- and $K^{\prime}$-band magnitudes \citep{vks02} may trace the mass in older stellar populations.  However, we have not corrected the latter for the contribution from supergiants, which for ULIGs could be substantial because of their intense nuclear star formation, or for the low-wavelength tail of far-infrared dust emission, which is sometimes non-negligible at these wavelengths.

ULIGs may be disky ellipticals-in-formation, and those classified as ellipticals based on their surface brightness profiles appear to lie on the fundamental plane \citep{genzel01,vks02}.  In method 2, we use values of the effective radii, $r_e$, and surface brightnesses, $\mu_e$, measured in the $R$ band \citep{vks02} to derive the central line-of-sight velocity dispersion $\sigma_0$ for each galaxy from the $R$-band fundamental plane \citep*{jfk96,hud97}.  The line-of-sight velocity dispersion is often used to represent random stellar motions, although it observationally depends on the aperture used to measure it and could include a rotational component.  Due to the intrinsic scatter on the fundamental plane and measurement uncertainties in $r_e$ and $\mu_e$, these velocity dispersions have an uncertainty of $\sim30\%$.

The emission and absorption lines used in method 3 are the following: in emission, [\ion{O}{3}] $\lambda\lambda4959,\;5007$; [\ion{O}{1}] $\lambda6300$; [\ion{N}{2}] $\lambda\lambda6548,\;6583$; H$\alpha$; and [\ion{S}{2}] $\lambda\lambda6716,\;6731$; and in absorption, the \ion{Ca}{2} triplet.  We also measured the line widths of $\mathrm{H}\gamma-\mathrm{H}11$ (excluding H$\epsilon$, which is mixed with \ion{Ca}{2} H) and \ion{Ca}{2} K in absorption, although we do not use these in method 3.  In most cases, the emission-line component of the lower order Balmer lines ($\mathrm{H}\beta-\mathrm{H}\delta$) was too prominent to yield accurate absorption line widths.  We list the results in Tables \ref{elw} and \ref{alw}.  For most emission lines, we measured the peak and continuum levels of each line and used these numbers to compute the location of the half-maximum points.  This technique was necessary because of the asymmetric nature of many of the strong emission lines.  This was done without fitting the continuum around the lines, but the surrounding continuum is generally flat.  Most absorption lines were fit with a Gaussian using the ``splot'' tool in IRAF; the level and slope of the continuum were estimated by eye for each line.  The values listed in Tables \ref{elw} and \ref{alw} are corrected in quadrature for an instrumental resolution of 65 \kms.  Measurement uncertainties for these line widths are $\sim5-10\%$ for most emission lines at the $\sim2$-$3\;\sigma$ level and $\sim20\%$ for most absorption lines (the absorption line uncertainties are higher due to lower sensitivity toward the red and blue parts of the spectrum, uncertainty in continuum levels, and intervening sky lines at long wavelengths).

We do not use the Balmer lines or \ion{Ca}{2} K in method 3 because the Balmer line widths are probably dominated by pressure-broadening rather than the stellar velocity dispersion of the galaxy, and the \ion{Ca}{2} K line contains a blend of stellar and interstellar absorption.  It is possible that the emission line widths are due to non-gravitational or non-circular motions, such as outflows or interactions; this may be especially true in AGNs for the [\ion{O}{3}] lines \citep{veil95,kvs98}.  It is also likely that different emission lines trace different parts of the galactic potential due to differences in lower levels, ionization potential, and critical density for de-excitation.  Taking an average over many emission lines may mitigate these effects; stellar absorption line widths should be largely free of them, however.

The tilt of the H$\alpha$ line in position-velocity space for several objects is suggestive of a rotation curve.  For three of these objects, we are able to estimate a lower limit to the amplitude of the rotation curve using the velocity difference between emission peaks (method 4).  These measurements are lower limits due to both unknown inclination effects and the fact that the emission peaks may not represent the full amplitude of the rotation curve (but rather only bright nuclear material).  Other objects have tilts, as well, but are too spatially unresolved or have no obvious peaks.  Some objects also show evidence for faint, spatially-extended, kinematically-disturbed line-emitting material.

The average emission line FWHM correlates well with the FWHM of the \ion{Ca}{2} triplet; see Figure \ref{elw_v_alw}.  Fitting a linear model that passes through 0 \kms, we arrive at
\begin{equation}
\mathrm{FWHM}_{\mathrm{avg}}^{\mathrm{em}}=(1.02\pm0.06) \times \mathrm{FWHM}_{\mathrm{avg}}^{\mathrm{CaII}},
\end{equation}
with a correlation coefficient of 0.99.  This may imply that the gaseous material and stellar populations traced by these lines have similar kinematics.  Furthermore, the measured kinematic quantities roughly correlate in the expected way with the measured luminosities based on the Tully-Fischer and Faber-Jackson relations; that is, more luminous (and more massive) objects have larger values of FWHM$_{\mathrm{avg}}^{\mathrm{em}}$, FWHM$_{\mathrm{avg}}^{\mathrm{CaII}}$, and $\sigma$.  The best correlation is seen in comparing $M_{K^{\prime}}$ and FWHM$_{\mathrm{avg}}^{\mathrm{em}}$; see Figure \ref{lw_v_mk}.

\subsubsection{Reheating efficiency as a function of galaxy mass}

In order to quantify the dependence of the reheating efficiency on the mass of the host galaxy, we plot it as a function of the measured kinematic quantities and absolute magnitudes in Figure \ref{re_v_all}.  These plots show no clear indication of a relationship between mass and reheating efficiency, even when we consider only objects that are dominated by star formation (objects that are optically classified as \ion{H}{2}-region-like or LINERs).  This is consistent with the results of \citet{mar99}, who finds no correlation between $\eta$ and maximum \ion{H}{1} rotation speed $v_c$ for her sample of less massive galaxies.  However, unresolved inclination effects could produce part of the scatter in the plot of $\eta$ vs. $\onehalf \mathrm{FWHM_{avg}}$, if the latter traces rotation.

\subsubsection{Escape fraction and ejection efficiency}

A related question is how much of the outflow has enough kinetic energy to escape the gravitational potential of its host.  To estimate the escape velocity for each galaxy, we assume a singular isothermal sphere potential.  For a singular isothermal sphere truncated at $r_{\mathrm{max}}$, the escape velocity $v_\mathrm{esc}$ at radius $r$ is related to the rotation speed $v_{c}$ and $r_{\mathrm{max}}$ by 
\begin{equation}
v_\mathrm{esc}(r) = \sqrt{2} v_{c} [1 + \ln{(r_{\mathrm{max}}/r)}]^{1/2}.
\end{equation}
We make a conservative estimate of $v_c$ by choosing the largest of \onehalf FWHM$_{\mathrm{avg}}^{\mathrm{em}}$, \onehalf FWHM$_{\mathrm{avg}}^{\mathrm{CaII}}$, $\sqrt{2}\sigma_0$, the rotation speed measured in position-velocity space, and, in the case of F$10190+1322$, the measured rotation speeds.  This makes the assumption that the FWHMs are largely due to bulk rotational (rather than random) motions, and that $\sigma_0$ represents random motions (the factor of $\sqrt{2}$ comes from equating the kinetic energy in bulk rotational and random motions).  Given that dark matter halos for these objects may extend to $\sim 100\;\mathrm{kpc}$, $r_{\mathrm{max}}/r$ is likely to be in the range $10-100$.  We choose $r_{\mathrm{max}}/r = 10$; although this means lower values for $v_\mathrm{esc}$, it is fairly insensitive to $r_{\mathrm{max}}/r$ anyway (increasing $r_{\mathrm{max}}/r$ by a factor of 10, from 10 to 100, only changes $v_\mathrm{esc}$ by $30\%$).  Our results on the escaping mass fraction also do not depend sensitively on the choice of $r_{\mathrm{max}}/r$, as we demonstrate below.

The escaping mass fraction in each object (Table \ref{cprop}) is defined as
\[
f_\mathrm{esc} \equiv \frac{\dot{M}_\mathrm{esc}}{\dot{M}_{\mathrm{tot}}}.
\]
To get $\dot{M}_\mathrm{esc}$, we compute the fraction of the equivalent width in each component that lies above the value of $v_\mathrm{esc}$ determined for each host galaxy, and then sum over all components; this is approximately valid for our sample, since all components are optically thin or close to it.  In Figure \ref{dv_v_vc}, we show how $\Delta v \pm \onehalf \mathrm{FWHM}$ (representing the velocity centroid and spread of each component) compares with $v_c$.  Those components with part or all of their profiles above the escape velocity for a given $v_c$ and $r_{\mathrm{max}}/r$ will have $f_\mathrm{esc} > 0$.

We also define a quantity, similar to the reheating efficiency $\eta$, that parameterizes the ``ejection efficiency'' for each object, or the efficiency with which star formation is able to expel gas from the galaxy:
\[
\delta \equiv \frac{\dot{M}_\mathrm{esc}}{\mathrm{SFR}}.
\]
Note that $\delta = \eta f_\mathrm{esc}$.  We list $\delta$ for each object in Table \ref{cprop}.  In their study of the mass-metallicity relation of ellipticals, \citet{kc98} use a prescription in which $\delta \sim v_c^{-2}$.  Although our sample contains only five objects with $\delta > 0$, we find no evidence for this dependence of $\delta$ on $v_c$ (or the other kinematic quantities we measure).

We find that $f_\mathrm{esc} > 0$ (and thus $\delta > 0$) for 6 out of 12 outflowing components and 5 out of 11 objects (disregarding the one component with $f_\mathrm{esc} = 0.01$).  Rather than averaging together individual values of $f_\mathrm{esc}$ and $\delta$, better measures of the global escape fraction and ejection efficiency for ULIGs at $z < 0.3$ are
\begin{eqnarray}
\langle f_\mathrm{esc} \rangle &=& \frac{\sum_{i=1}^{11} \dot{M}_\mathrm{esc}^i}{\sum_{i=1}^{11} \dot{M}_{\mathrm{tot}}^i} = 0.49 \\
\mathrm{and}\;\langle \delta \rangle &=& \frac{\sum_{i=1}^{11} \dot{M}_\mathrm{esc}^i}{\sum_{i=1}^{11} \mathrm{SFR}^i} = 0.10
\end{eqnarray}
where the sums are taken over all objects but Mrk 231.

These two results are relatively insensitive to the choice of $r_{\mathrm{max}}/r$ and the fact that we assumed the FWHMs to be due to rotation, rather than random motions.  In the case where we assume $r_{\mathrm{max}}/r = 100$, the result is $\langle f_\mathrm{esc} \rangle = 0.41$ and $\mathrm{and}\;\langle \delta \rangle = 0.09$.  In the case where we assume that the FWHMs trace random motions (i.e. $v_c = \sqrt{2} \onehalf \mathrm{FWHM}$), $\langle f_\mathrm{esc} \rangle = 0.40$ and $\mathrm{and}\;\langle \delta \rangle = 0.08$.  Assuming both results in $\langle f_\mathrm{esc} \rangle = 0.33$ and $\langle f_\mathrm{esc} \rangle = 0.07$.  However, as we increase $v_\mathrm{esc}$, the number of objects for which $f_\mathrm{esc}$ is non-negligible (i.e. greater than a few percent) decreases, such that there are only two remaining objects in the extreme case of assuming large $r_{\mathrm{max}}/r$ and random FHWMs.

These results suggest that $\sim40-50\%$ of the total outflowing gas in all ULIGs at $z < 0.3$ escapes into the intergalactic medium, and that star formation is relatively efficient at expelling gas from these galaxies, ejecting $\sim8-10\%$ of the mass of interstellar gas that goes into star formation.  However, it should be emphasized that these values are fairly uncertain.  We do not know how the radius at which we have computed $v_\mathrm{esc}$ matches the actual radius of the absorbing gas; we do not have the data to model the potential of these galaxies in detail; and small number statistics or selection biases could play a role in our measurements.  Note also that unknown inclination effects could affect our calculation of $v_c$ and $v_\mathrm{esc}$; however, as we discuss above, $\langle f_\mathrm{esc} \rangle$ is relatively insensitive to our choice of $v_\mathrm{esc}$.  Finally, and perhaps most importantly, other forces (e.g. hydrodynamic and magnetic) besides the gravitational force are at work to accelerate or slow the wind \citep{strick02}.  For instance, the wind also might end up transferring energy to dense clouds of gas or a dense halo, lose energy in turbulent vortices, or be contained by magnetic fields, in which case the escape fraction is in reality smaller than we measure.

The escape fraction of hot wind material (to which we are insensitive) may be higher, due to its greater specific energy (measured against the binding energy of the galaxy).  Hot gas that is interior to a shell of cool, swept-up ambient material in an expanding superbubble is able to exit the bubble when the shell ruptures and fragments.  Simulations of dwarf galaxies with low star formation rates suggest that the percentage of this hot material that escapes the galaxy is greater than the percentage of cooler, entrained gas that escapes; further, galaxies with low masses are the most efficient at ejecting this material \citep{maclow99}.  These results are not immediately applicable to our sample, since the mechanical energy injection rates from star formation would have to be scaled upward substantially and \citet{maclow99} do not include the possibility of a surrounding dense halo or complex of tidal debris (which may be present around ULIGs if they are created by galaxy interactions).  \citet{st01} perform a similar study of dwarf galaxies, including the effects of halos and external pressure from the surrounding IGM, and conclude that halos substantially increase the threshold energy required to eject gas into the IGM.  In general, more gas than we detect in this survey may actually escape these galaxies unless the large masses of ULIGs or surrounding gas becomes an important factor.

\subsubsection{Cosmological implications}

What implications do our measurements have for the fractional contribution of ULIGs to the total amount of gas escaping galaxies and entering the intergalactic medium as a function of $z$?  It is straightforward to show that this fraction is given by
\begin{equation}
Q(z) = P(z)\;\frac{1}{P(z) + \langle \delta \rangle^{\mathrm{sfg}}(z) / \langle \delta \rangle^{\mathrm{ULIGs}}(z) [1 - P(z)]},
\end{equation}
where $P(z)$ is the fractional contribution of ULIGs to the total star formation rate density of the universe at redshift $z$ and $\langle \delta \rangle^{\mathrm{sfg}}(z)$ is the ejection efficiency for all star-forming galaxies other than ULIGs as a function of $z$.  In Figure \ref{q_v_dd}, we plot $Q$ as a function of $\langle \delta \rangle^{\mathrm{sfg}} / \langle \delta \rangle^{\mathrm{ULIGs}}$ for various values of $P$.

Locally (i.e. our sample at $z < 0.3$), the fraction of total star formation that occurs in ULIGs is modest ($\sim 5\%$; \citealt{adel01}).  However, because of the strong increase in the number density of ULIGs with $z$ (up to at least $z \sim 2-3$; see references in \S \ref{intro}), this fraction may increase strongly, as well.  If $P(z)$ continues to increase with $z$, ULIGs at higher $z$ may contribute a significant fraction of the gas being expelled into the IGM.  If we assume that $\langle \delta \rangle^{\mathrm{sfg}} / \langle \delta \rangle^{\mathrm{ULIGs}} \sim 10$ at all $z$ (i.e. equal to the value we measure locally), then $P(z)$ must be greater than $50\%$ to get $Q(z) \ga 10\%$.  In other words, ULIGs must produce more than half of all star formation at a given $z$ to be responsible for more than $10\%$ of the mass escaping galaxies and entering the IGM at that $z$.

\section{CONCLUSIONS}

We present the results of a moderately-high resolution spectroscopic study of a sample of 11 ULIGs.  These objects are selected from the IRAS 1 Jy sample of \citet{ks98} as good candidates to contain massive, galaxy-scale outflows.  (This possible bias toward objects with high mass outflow rates should be considered when assessing the generality of our results.)  The goal of the study is to use prominent absorption lines to measure mass outflow rates in these galaxies.  Profiles are fit to the \ion{Na}{1} D lines in each object to measure column densities, $N$(\ion{Na}{1}); velocities relative to the host galaxy, $\Delta v = v_{\mathrm{sys}} - v$; line widths; and a covering fraction $C_f$.  Our analysis technique assumes a Gaussian velocity distribution for the absorbing gas and fits intensity profiles using a $\chi^2$ minimization; it is useful for fitting profiles with one or two absorbing components (or more if well-separated in velocity space or otherwise constrained), profiles with a covering fraction less than unity, and even slightly saturated profiles.

We measure absorption-line components blueshifted by $\ga 70\;\kmsm$ in 8 of the 11 targets ($73\%$); this stands in between the recent measurements of HLSA and \citet{pet01}, who find outflows in $38\%$ and $100\%$ of their sample, respectively.  The absorbing material in our sample have relatively low optical depths (mostly $\tau_{1,c} < 1.0$), and the corresponding covering fractions span a wide range.  The typical maximum outflow velocity $\Delta v_{\mathrm{max}}$ and total outflowing \ion{Na}{1} column density are $\sim 300\;\kmsm$ and $(4-5)\times10^{13}\;\mathrm{cm}^{-2}$, respectively (although one object has $\Delta v_{\mathrm{max}} = 1540\;\kmsm$).  We also measure the host galaxy kinematics in several ways and present absolute $R$- and $K^{\prime}$-band magnitudes \citep{vks02} in order to trace the masses and gravitational potentials of these galaxies.  From these data, we infer the following:

(1) {\it Mass outflow rate.}  Using a simple model of a mass-conserving free wind and assuming that all absorbing components with $\Delta v > 70\;\kmsm$ are outflowing, the corresponding total mass outflow rates for objects hosting a wind are in the range $13-133\;\smpym$.  The simplicity of our model and the difficulty of ionization corrections in the conversion from $N$(\ion{Na}{1}) to $N$(H) likely dominate the uncertainties in $\dot{M}$.  However, the absorbing gas is probably in harsher conditions than we assume, suggesting that our results are lower limits to the actual values.

(2) {\it Reheating efficiency.}  The reheating efficiencies in our sample, equal to $\dot{M}_{\mathrm{tot}}$ divided by the corresponding global star formation rate ($\eta \equiv \dot{M}_\mathrm{tot} / \mathrm{SFR}$), are in the range $0.1-0.7$ for galaxies hosting a wind.  These values are on average a factor of only $\sim10$ smaller than those measured by \citet{mar99}, who studied warm ionized and hot gas in a sample of nearby dwarf galaxies and edge-on spirals and concluded $\eta = 0.7-5$, and \citet{heck00} (HLSA), who studied a sample of luminous and ultraluminous infrared galaxies and concluded that $\eta \sim 1$.  Given that the galaxies in these studies have star formation rates that are, on average, $10-10^4$ times smaller than the SFRs in ULIGs, and that molecular gas in ULIGs may inhibit outflows, it is remarkable that their values for $\eta$ differ by a factor of only 10.  This implies that the outflow rate of a wind scales (to within a factor of 10) with the corresponding star formation rate over a wide range of values of SFR, which in turn suggests that the physical conditions governing outflows may be somewhat similar in both ULIGs and less massive galaxies.

(3) {\it $\eta$ and $\delta$ vs. host mass.}  Within our sample, the reheating efficiency apparently has no dependence on the mass of the host galaxy (as traced by emission- and absorption-line widths, central velocity dispersions, $M_R$, and $M_{K^{\prime}}$), a result also obtained by \citet{mar99}.  Although our subsample of objects with ejection efficiency $\delta \equiv \dot{M}_\mathrm{esc} / \mathrm{SFR} > 0$ is small ($\sim 5$ objects), we also find no evidence that $\delta \sim v_c^{-2}$, a prescription postulated by \citet{kc98} in their theoretical study of the mass-metalliticy relation of ellipticals.

(4) {\it Escape fraction and ejection efficiency.}  We estimate the fraction of absorbing gas in our galaxies which escapes into the surrounding medium by using host galaxy kinematics and a singular isothermal sphere potential to compute $v_\mathrm{esc}$ and compare it to the outflowing gas velocities.  We find that 5 out of 11 objects have non-zero escape fractions, and the resulting average escape fraction of gas is $\langle f_\mathrm{esc} \rangle \equiv \sum \dot{M}_\mathrm{esc}^i / \sum \dot{M}_{\mathrm{tot}}^i = 0.4-0.5$.  The corresponding average ``ejection efficiency'' is $\langle\delta\rangle \equiv \sum \dot{M}_\mathrm{esc}^i / \sum \mathrm{SFR}^i = 0.08-0.10$.  These values are fairly uncertain, and could be compromised by any sample selection bias or the small size of our sample, but are relatively insensitive to the way in which we calculate $v_\mathrm{esc}$.

At a given $z$, the fraction of gas being expelled into the intergalactic medium that originates in ULIGs is governed by two parameters: (1) the fractional contribution of ULIGs to the total star formation rate density of the universe, and (2) the ratio of the average ejection efficiency of other star-forming galaxies to that of ULIGs, $\langle \delta \rangle^{\mathrm{sfg}} / \langle \delta \rangle^{\mathrm{ULIGs}}$.  The fractional contribution of ULIGs to star formation is modest in the local universe, but it may increase strongly with $z$ due to the observed strong increase in the number density of ULIGs with $z$.  For $\langle \delta \rangle^{\mathrm{sfg}} / \langle \delta \rangle^{\mathrm{ULIGs}} = 10$ (consistent with current observations), we find that ULIGs must host at least $50\%$ of all star formation at a given $z$ for the fractional contribution of ULIGs to gas ejected from galaxies into the IGM to be greater than $10\%$.

\acknowledgments

The authors thank the referee, Tim Heckman, for a thoughtful and detailed referee report.  DSR thanks Ray Weymann and Cole Miller for reading versions of the manuscript and helpful discussions and comments.  SV is grateful for partial support of this research by a Cottrell Scholarship awarded by the Research Corporation, NASA/LTSA grant NAG 56547, and NSF/CAREER grant AST-9874973.  DSR is also supported by NSF/CAREER grant AST-9874973.  This research has made use of the NASA/IPAC Extragalactic Database (NED), which is operated by the Jet Propulsion Laboratory, California Institute of Technology, under contract with the National Aeronautics and Space Administration.

\appendix

\section{MARKARIAN 231} \label{append}

Mrk 231 (listed as F$12540+5708$ in Table \ref{sprop}) is the only optically classified Seyfert 1 galaxy in our sample.  This object contains broad, complex absorption line systems in low-ionization species, including \ion{Na}{1} D, \ion{He}{2} $\lambda3889$, and \ion{Ca}{2} H and K \citep*{bok77,rfs85,hutch87,boroson91,frm95}.  Some components of these absorption line systems have proven to be time-variable \citep{hutch87,boroson91,frm95}.  The dramatically different nature of these absorption lines compared to those in the ULIGs discussed above (i.e. broader, blacker, and higher velocity), coupled with the fact that Mrk 231 is a Seyfert 1, suggest that their physical origin is distinct.

In Figure \ref{mrk231}, we present our observations of and fits to the absorption lines discussed above.  Note that \ion{Na}{1} D is also present as a broad emission line near systemic, although this is not visible in Figure \ref{mrk231}.  Using the analysis of \citet{frm95} as a starting point, we fit the \ion{Na}{1} D lines in this object using 12 narrow components with relatively small optical depths, despite the initial appearance of strong saturation in the profile.  Our results are comparable to those of \citet{frm95}, although we also allowed for a partial covering fraction in each component.  Using the velocity components measured in absorption in \ion{Na}{1}, we fit the strong absorption features of \ion{He}{1} $\lambda 3889$ and \ion{Ca}{2} H \& K.  We also constrained the covering fraction in each component to be the same as for \ion{Na}{1} (and in a few cases, the FWHM of the component, as well, to avoid broad lines in the fit).  The resulting velocity components and values for $C_f$, $\tau$, $b$, and $N$ are listed in Table \ref{table_mrk231}.  It is important to note that this solution is not very well constrained or unique, given the number of components used to fit each line, suggesting that the parameter values measured in the blended parts of the lines are fairly uncertain.  There is also no {\it a priori} reason to expect that $C_f$ will be the same for different ions, but we make this assumption to provide better constraints on the fit.

In instances where two or more lines from a given ion differing in the product $f \lambda$ are present, a more physically sound way to perform this anaylsis is to compute the optical depth and covering fraction as a function of velocity.  Because the \ion{Na}{1} D doublet lines are closely spaced, this treatment is not possible; however, it is feasible for the \ion{Ca}{2} H \& K lines, which have a large spacing and an $f \lambda$ (and optical depth) ratio of 2.  In Figure \ref{mrk231caii}, we present these Ca lines as a function of $\Delta v$; we have rebinned the spectra into $log(\lambda)$ space, binned again by 6 pixels (corresponding to 68 \kms, the resolution of our spectra), and boxcar smoothed by 3 bins.  The covering fraction and optical depth in each velocity bin, shown as $(1-C_f)$ and $e^{-2 \tau_1}$ in Figure \ref{mrk231}, are then given by \citep{hamann97}
\begin{eqnarray}
C_f &=& \left\{	\begin{array}{ccl}
	\frac{I_1^2 - 2 I_1 + 1}{I_2 - 2 I_1 + 1} &\mathrm{if}& I_1 > I_2 \geq I_1^2 \\
	1 &\mathrm{if}& I_2 < I_1^2 \\
	1 - I_1 &\mathrm{if}& I_2 \geq I_1
		\end{array}\right. \\
\tau_1 &=& \mathrm{ln} \left(\frac{C_f}{I_1 + C_f - 1}\right) \\
\mathrm{and}\;\;\tau_2 &=& 2 \tau_1,
\end{eqnarray}
where $I_1$ ($\tau_1$) and $I_2$ ($\tau_2$) are the residual intensities (optical depths) in \ion{Ca}{2} H $\lambda3968$ and \ion{Ca}{2} K $\lambda3934$, respectively.  It appears that the covering fraction in this ionization state of Ca is smaller at higher optical depths (i.e. near line center), and that the intensity at higher optical depths is determined by $C_f$, rather than $\tau$.  This result is similar to that found by \citet{arav99b} for the \ion{Si}{4} $\lambda\lambda 1394,\;1403$ doublet in the BAL QSO $1603+3002$, where the line profile becomes optically thick but not black and the residual intensity in the lines is determined by $C_f$.  It also appears that our multi-component fits to \ion{Ca}{2} reproduce $\tau_2$ and $C_f$ computed using the above equations in some cases, but not all (see the points in Figure \ref{mrk231caii}).  This may also be the case in the \ion{He}{1} and \ion{Na}{1} lines in our observations, but it is impossible to distinguish the effects of $\tau$ and $C_f$ as a function of velocity without more information (i.e. other transitions).

The detailed modelling and interpretation of these results is outside the scope of this paper.  However, we do compute a mass outflow rate for Mrk 231 based on our \ion{Na}{1} fits.  We measure a total outflowing column density of $N(\mathrm{Na\;\mbox{\small I}})=3.1\times10^{14}\;\mathrm{cm}^{-2}$, which is slightly higher than the $1.9\times10^{14}\;\mathrm{cm}^{-2}$ measured by \citet{frm95}.  Normalizing to $r_{\star} = 1\;\mathrm{pc}$, we get
\begin{equation}
\dot{M} = 5.2\;\left(\frac{r_{\star}}{1\;\mathrm{pc}}\right)\;\smpym.
\end{equation}
We use a smaller value for $r_{\star}$ because the energy injection region for the outflow is likely to be smaller, especially given the variability observed in the source and the high velocities, which seem to indicate that the AGN is driving the outflow, rather than the starburst.  We do not compute a reheating efficiency for Mrk 231; if the AGN powers the wind, $\eta$ is probably not a meaningful quantity (since it involves the star formation rate of the host).

The absorption system at $\Delta v = 8060\;\kmsm$ is time-variable.  Previous non-detections and detections include the following: (1) 1984 December, \citet{hk87}, no detection in \ion{Na}{1}; (2) 1988 May, \citet{boroson91}, detection in \ion{Na}{1} ($\Delta v = 7760\;\kmsm$---after conversion to $z = 0.04217$---and $W_{\mathrm{eq}}^{\mathrm{rest}} = 2.1\;\mbox{\AA}$) and in \ion{He}{1} $\lambda 3889$; (3) 1991 April, \citet{frm95}, detection in \ion{Na}{1} D (7830 \kms, 1.7 \AA); (4) 1994 April, \citet{frm95}, detection in \ion{Na}{1} D (7860 \kms, 0.9 \AA).  We measure a (rest-frame) equivalent width of 0.36 \AA\ for this feature at $\lambda_1=5983.1\;\mbox{\AA}$, but we do not detect it in \ion{He}{1} or \ion{Ca}{2}.  (The sensitivity of the spectrograph is declining sharply at the expected location of the \ion{He}{1} feature, so the fact that we do not detect it could be due to lack of sensitivity rather than to the absence of the feature.)  It is interesting to note that the strength and blueshifted velocity of this feature in \ion{Na}{1} D have decreased and increased, respectively, over time.

\clearpage

\begin{figure}
\plotone{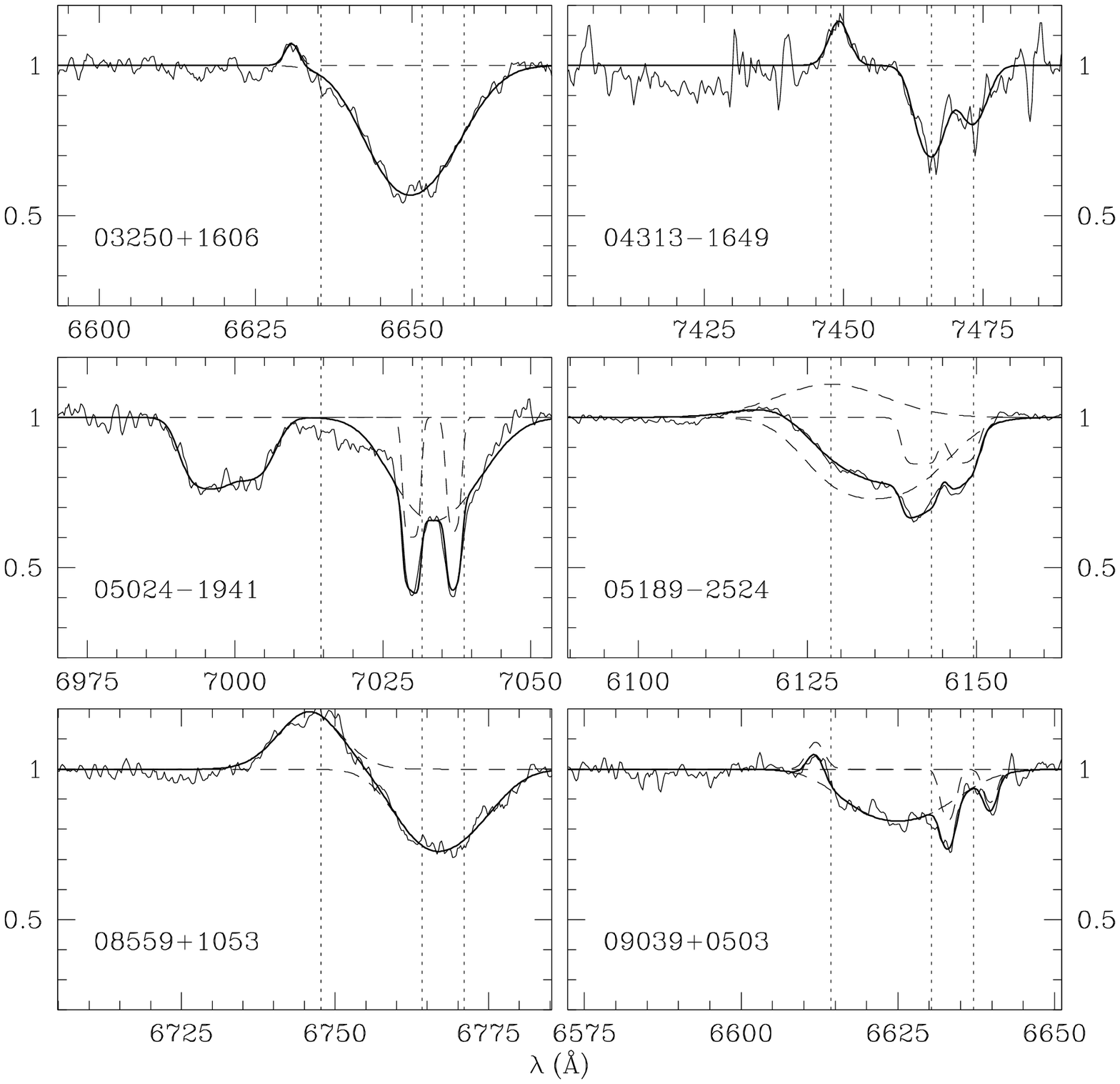}
\caption{Spectra of the \ion{Na}{1} D line in our 11 program objects.  The thin, jagged, solid lines are the original spectra smoothed with a boxcar of width 3 pixels, and the thick, smooth, solid lines are the fits to the data.  The dashed lines indicate the individual components of the fit.  The vertical dotted lines locate the \ion{Na}{1} D doublet and \ion{He}{1} $\lambda5876$ line in the rest frame of the galaxy.}
\label{spectra}
\end{figure}

\setcounter{figure}{0}
\begin{figure}
\epsscale{}
\plotone{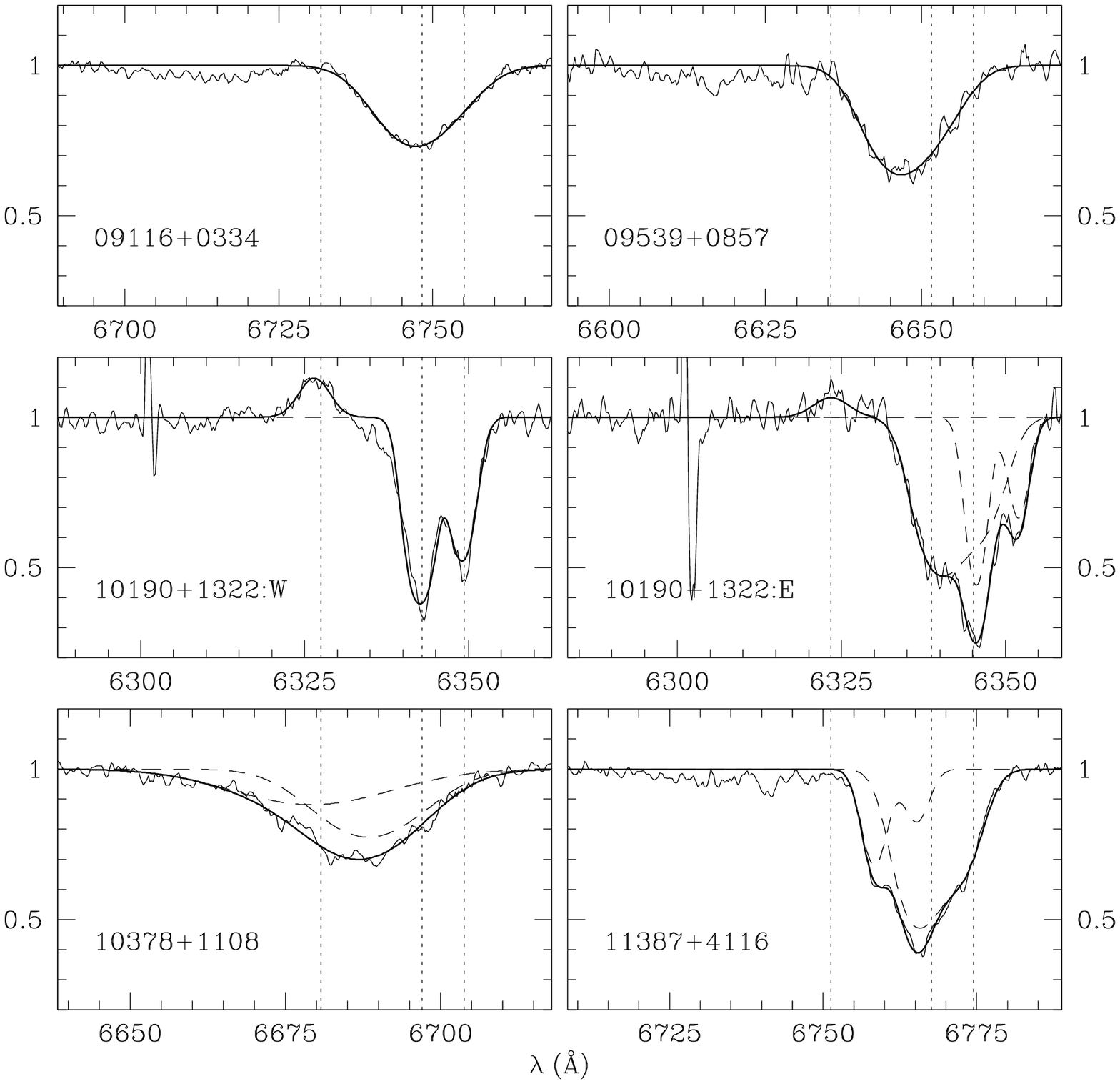}
\caption{\it{Continued.}}
\end{figure}

\begin{figure}
\plotone{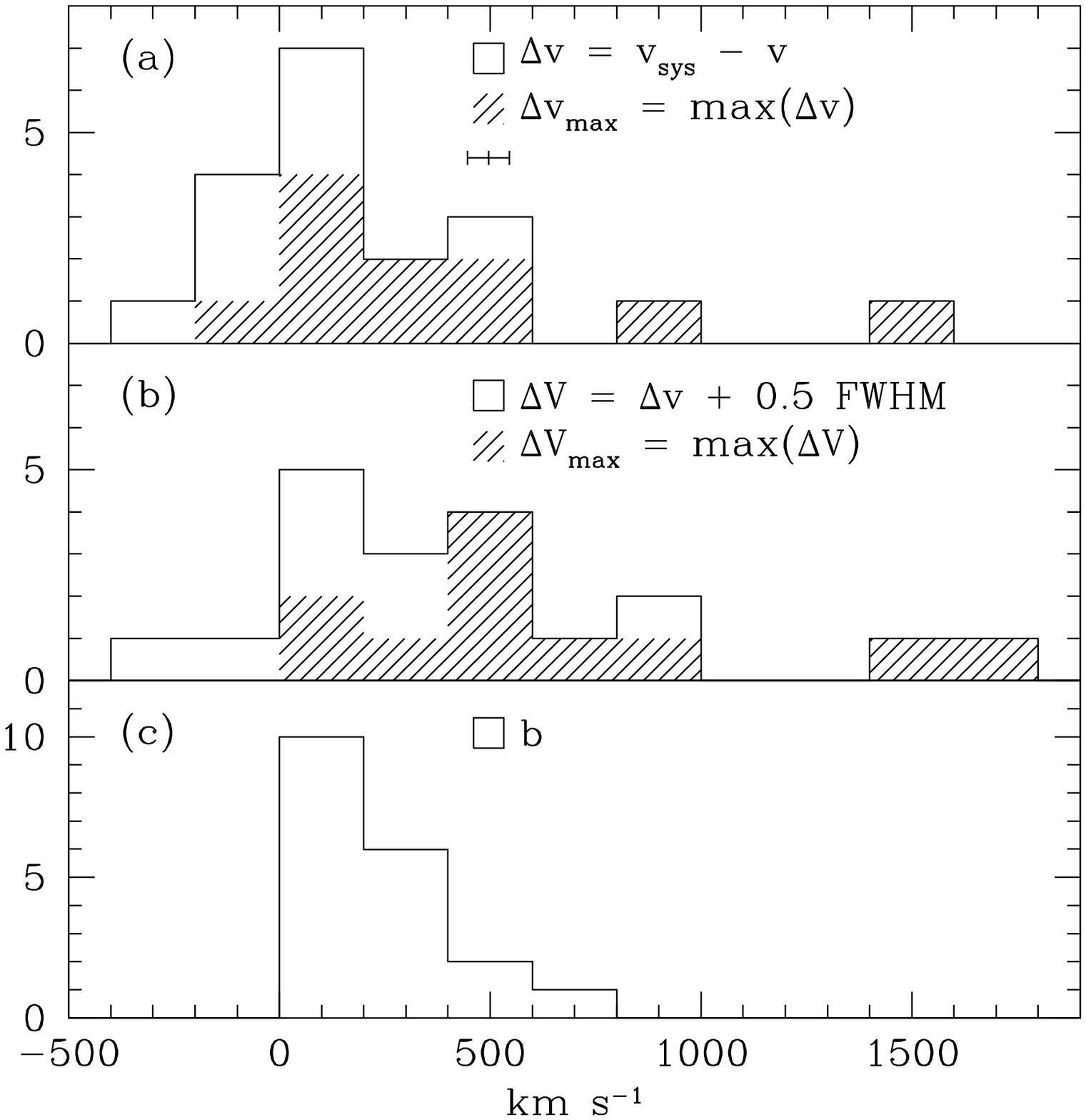}
\caption{Distributions of various kinematic quantities: (a) velocity relative to systemic for each component, $\Delta v = v_{\mathrm{sys}} - v$, unshaded; and maximum $\Delta v$ for each object, $\Delta v_{\mathrm{max}} = max(\Delta v)$, shaded; (b) highest velocity within each component, $\Delta V = \Delta v + \onehalf \mathrm{FWHM}$, unshaded; and maximum $\Delta V$ for each object, $\Delta V_{\mathrm{max}} = max(\Delta V)$, shaded; and (c) Doppler parameter $b$ for each component.  Note that blueshifted velocities are positive, and that FWHM (and thus $b$) is corrected in quadrature for a 65 \kms\ instrumental resolution.  The error bar in (a) represents a typical $\sim2$-$3\;\sigma$ uncertainty.}
\label{dvhist}
\end{figure}

\begin{figure}
\plotone{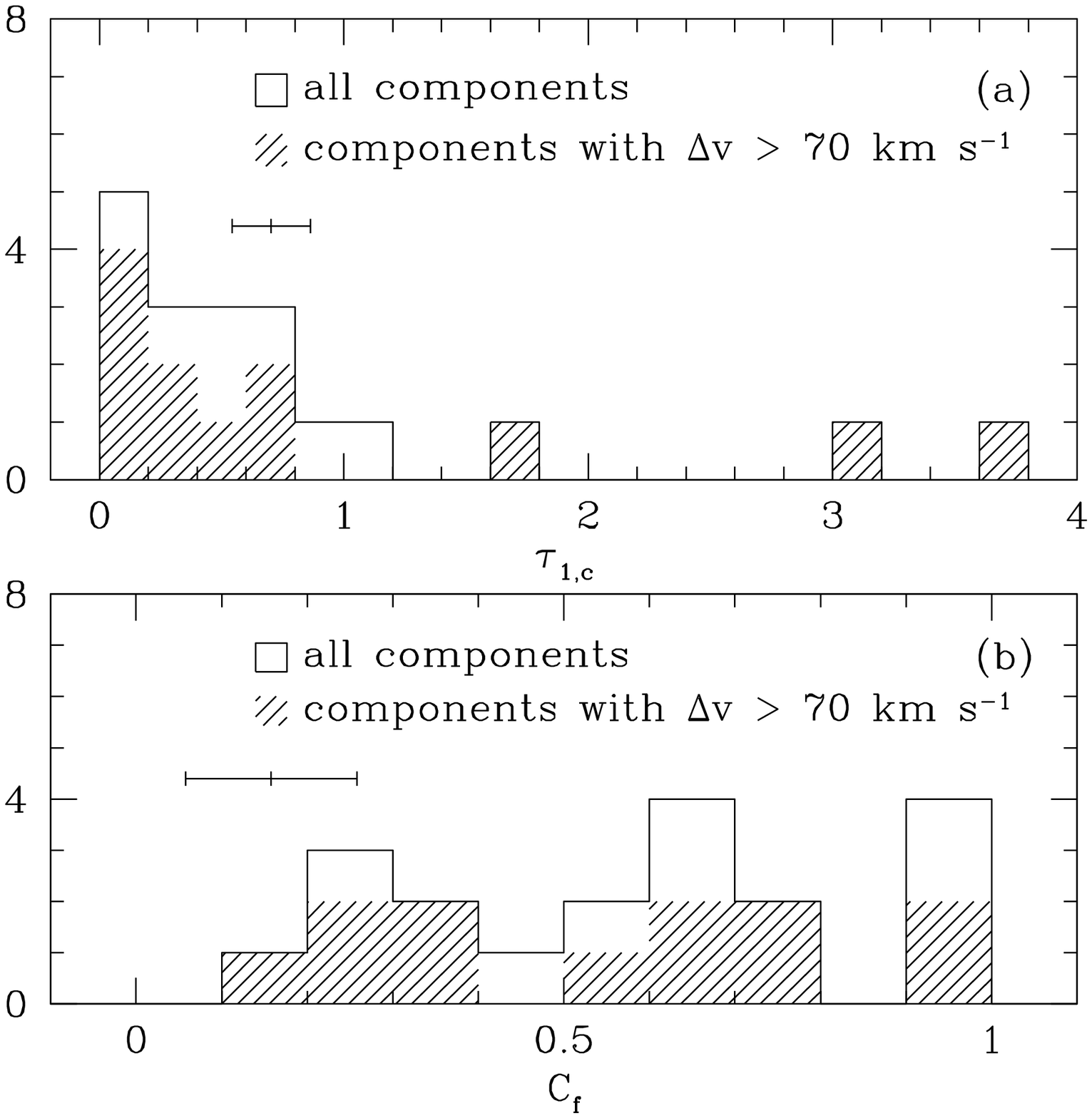}
\caption{Distributions of (a) central optical depth $\tau_{1,c}$ of the \ion{Na}{1} D$_1$ $\lambda5896$ line and (b) covering fraction $C_f$.  Unshaded histograms include all components; shaded histograms include only components with $\Delta v > 70\;\kmsm$ (i.e. those assumed to be outflowing).  The error bars represent typical $\sim2$-$3\;\sigma$ uncertainties.}
\label{cfhist}
\end{figure}

\begin{figure}
\plotone{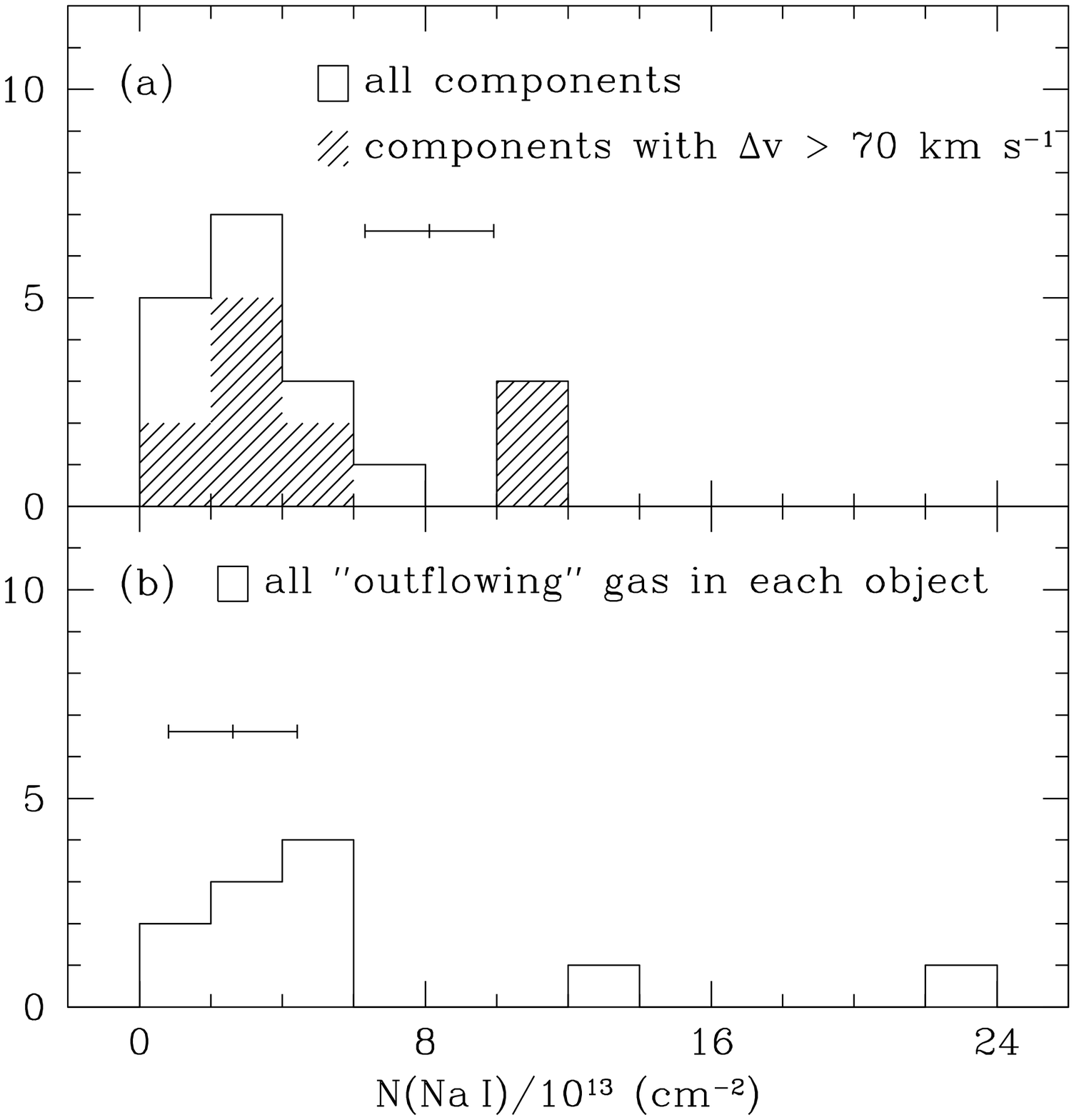}
\caption{Distributions of (a) column density of \ion{Na}{1} for each component (unshaded) and only components with $\Delta v > 70\;\kmsm$ (shaded) and (b) total column density of \ion{Na}{1} gas assumed to be outflowing in each object.  The error bars represent typical $\sim2$-$3\;\sigma$ uncertainties.}
\label{nhist}
\end{figure}

\begin{figure}
\plotone{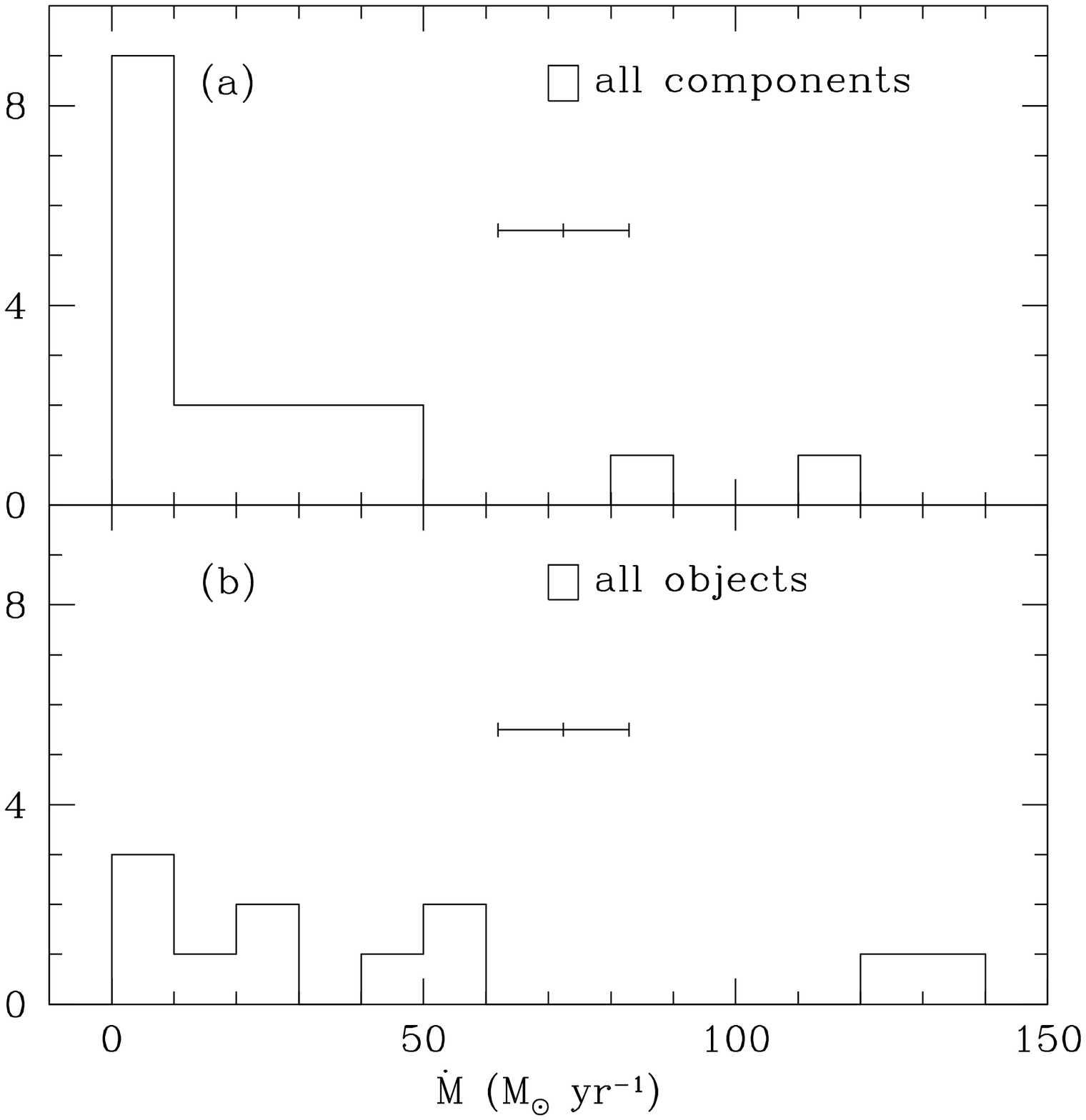}
\caption{Distributions of (a) mass outflow rate $\dot{M}$ for each component and (b) total mass outflow rate $\dot{M}_{\mathrm{tot}}$ for each object.  The error bars represent typical $\sim2$-$3\;\sigma$ uncertainties.}
\label{mhist}
\end{figure}

\begin{figure}
\plotone{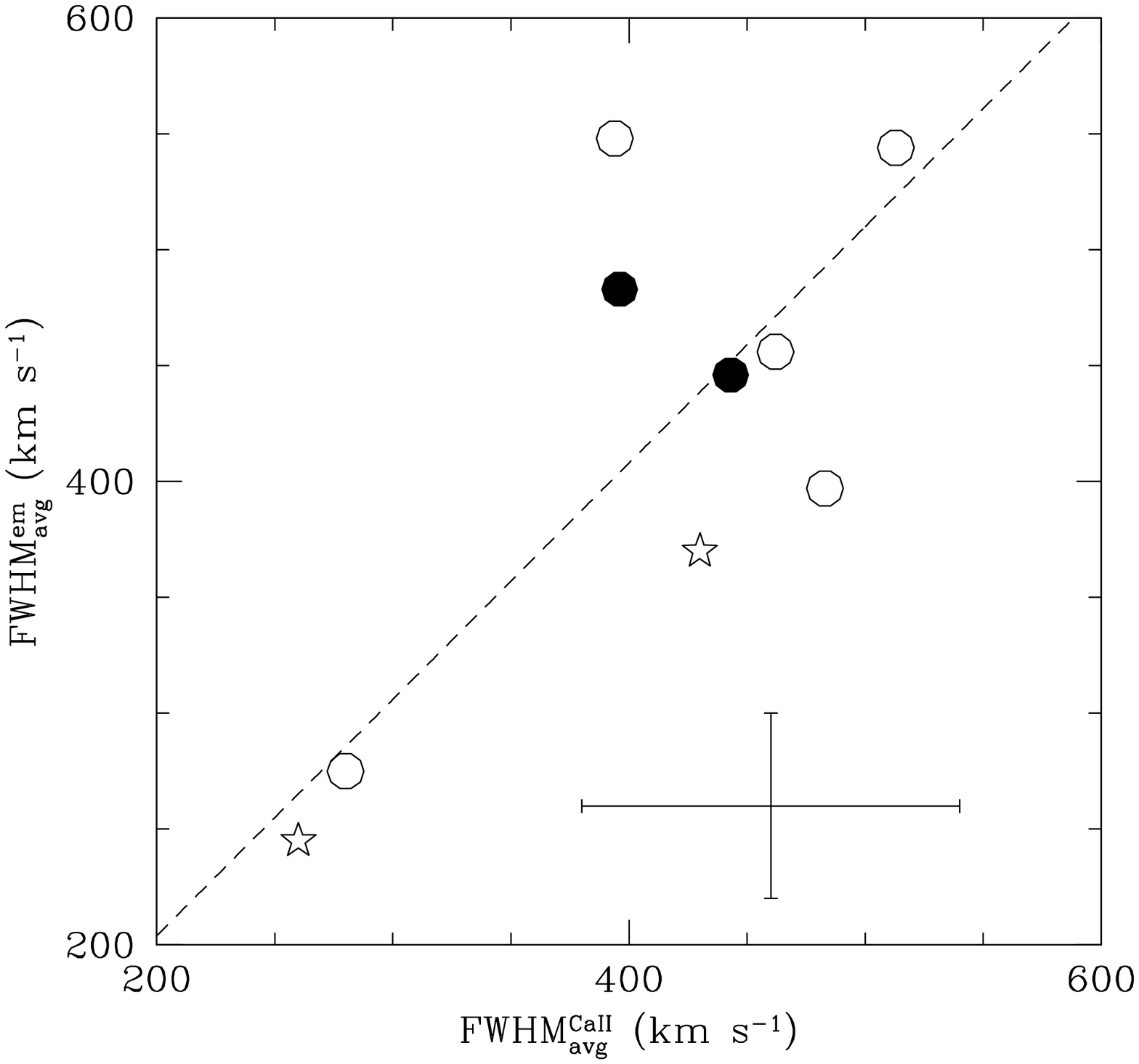}
\caption{Average emission line FWHM vs. average FWHM of \ion{Ca}{2} triplet absorption feature ($\lambda\lambda8498,\;8542,\;8662$).  The two quantities correlate; a linear regression yields $\mathrm{FWHM}_{\mathrm{avg}}^{\mathrm{em}}=(1.02\pm0.06) \times \mathrm{FWHM}_{\mathrm{avg}}^{\mathrm{CaII}}$, with a correlation coefficient of 0.99.  The dashed line is the fit.  Symbols denote optical spectral type: open star = \ion{H}{2}-region-like object; open circle =  LINER; filled circle = Seyfert 2.  The error bars represent typical $\sim2$-$3\;\sigma$ uncertainties.}
\label{elw_v_alw}
\end{figure}

\begin{figure}
\plotone{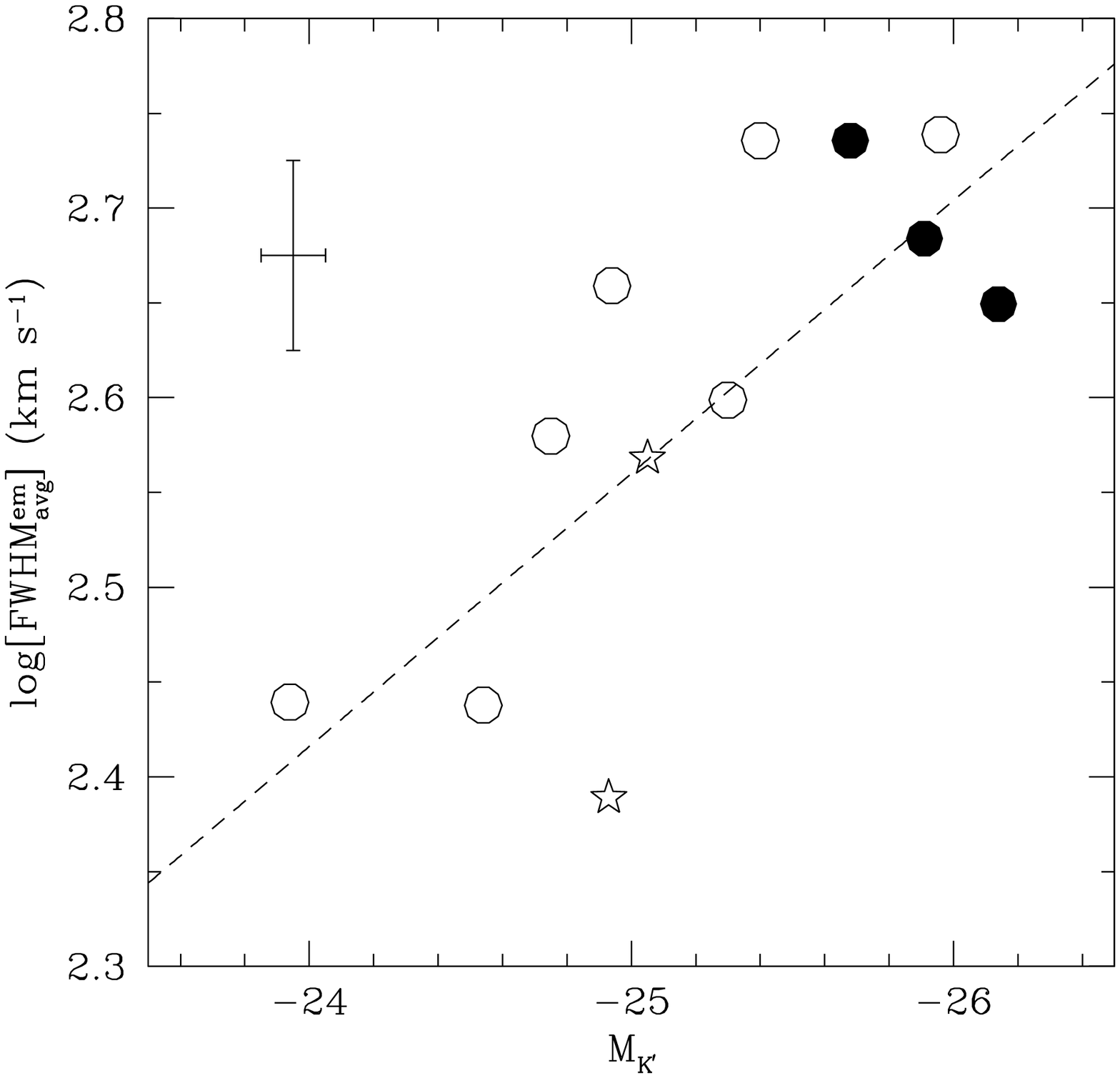}
\caption{Logarithm of average emission line FWHM vs. absolute $K^{\prime}$-band magnitude.  Dashed lines indicate the fit; a linear regression yields $log[\mathrm{FWHM}_{\mathrm{avg}}^{\mathrm{em}}]=(0.14\pm0.04) \; M_{K^{\prime}} - (1.04\pm0.98)$ with a correlation coefficient of 0.76.  As in Fig. \ref{elw_v_alw}, symbol shapes denote optical spectral type.  The error bars represent typical $\sim2$-$3\;\sigma$ uncertainties.}
\label{lw_v_mk}
\end{figure}

\begin{figure}
\plotone{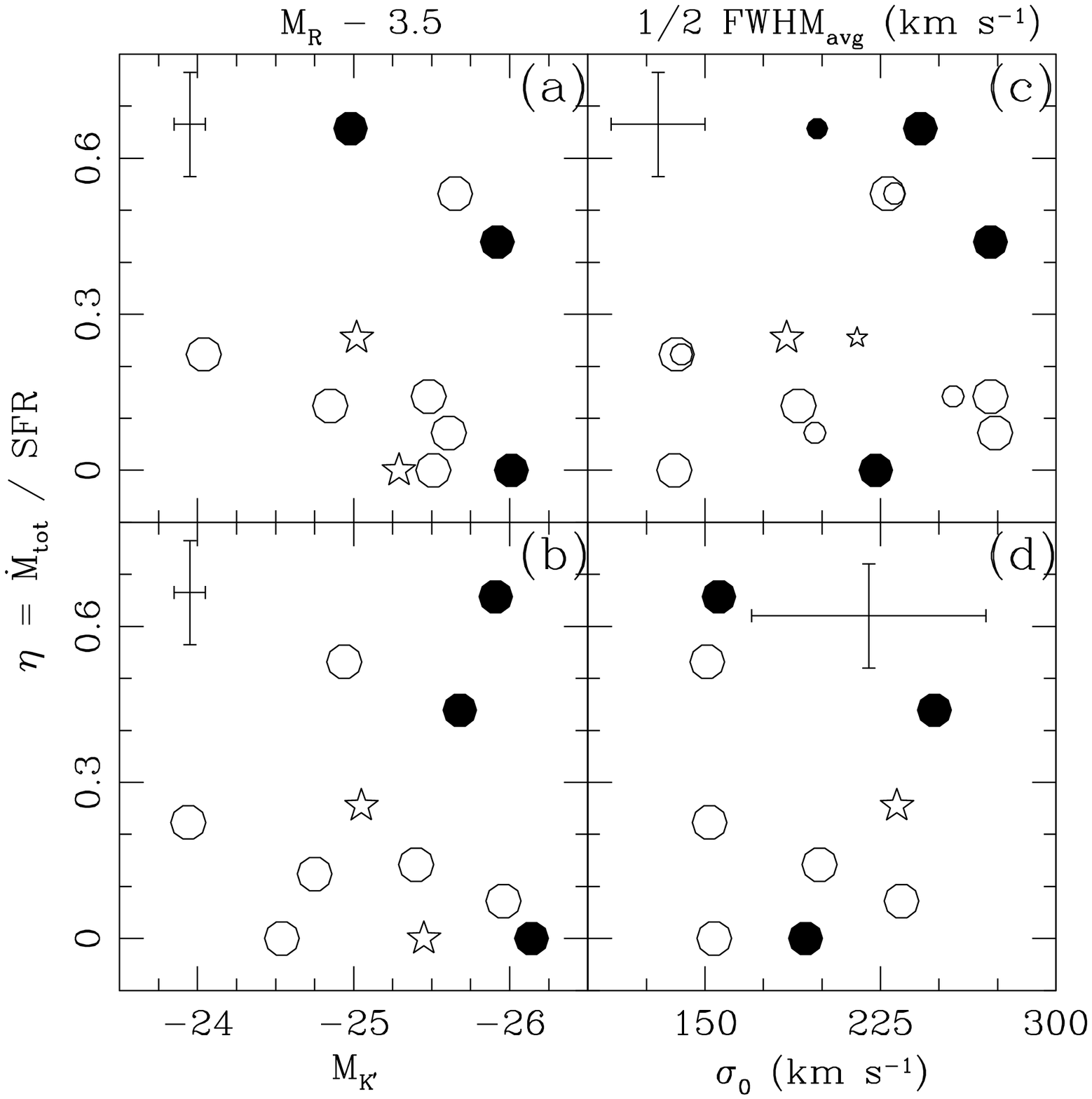}
\caption{Reheating efficiency $\eta = \dot{M} / \mathrm{SFR}$ vs. various quantities: (a) absolute $R$-band magnitude; (b) absolute $K^{\prime}$-band magnitude; (c) one-half the average emission-line FWHM (large symbols) and \ion{Ca}{2} triplet FWHM (small symbols); and (d) the central velocity dispersion computed from the fundamental plane.   There appear to be no correlations for the entire sample or any one spectral type.  As in previous figures, symbol shapes denote optical spectral type.  The error bars represent typical $\sim2$-$3\;\sigma$ uncertainties.}
\label{re_v_all}
\end{figure}

\begin{figure}
\plotone{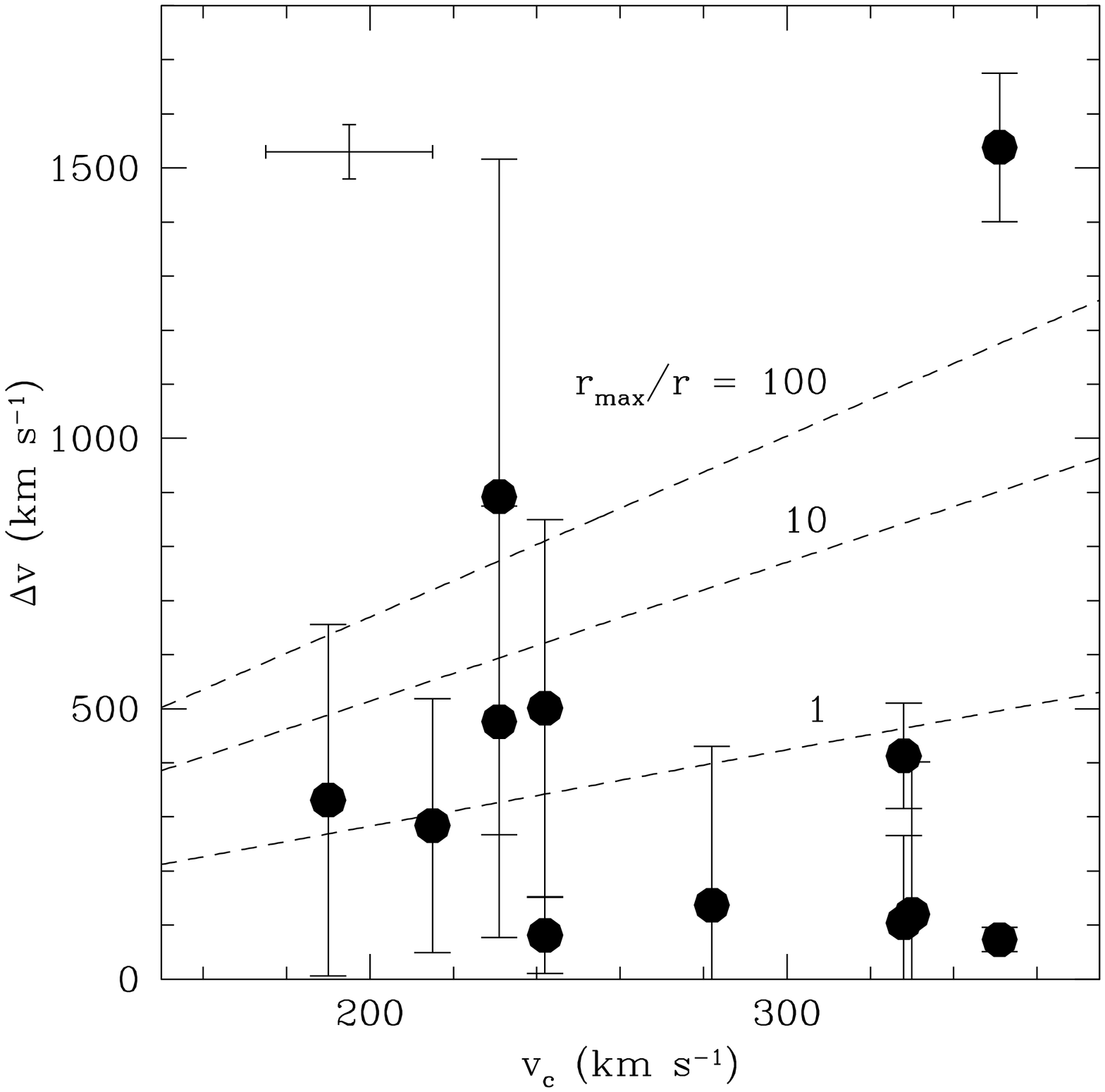}
\caption{Velocity relative to systemic of each component assumed to be outflowing vs. estimated circular velocity of host galaxy.  The vertical bar on each point represents the FWHM of the corresponding fitted profile in velocity space.  The dotted lines are values of $v_\mathrm{esc}$ for a singular isothermal sphere truncated at $r_{\mathrm{max}}$, for various values of $r_{\mathrm{max}}/r$.  Profiles that extend above a given value of $v_\mathrm{esc}$ may represent escaping material.  We adopt $r_{\mathrm{max}}/r = 10$ in the text, but the average escape fraction in the sample is not sensitive to this quantity.  The cross in the upper left-hand corner represents measurement uncertainties for $\Delta v$ and $\onehalf \mathrm{FWHM^{em}_{avg}}$; if $\onehalf \mathrm{FWHM^{CaII}_{avg}}$ or $\sigma_0$ are used to determine $v_c$, the errors are larger.}
\label{dv_v_vc}
\end{figure}

\begin{figure}
\plotone{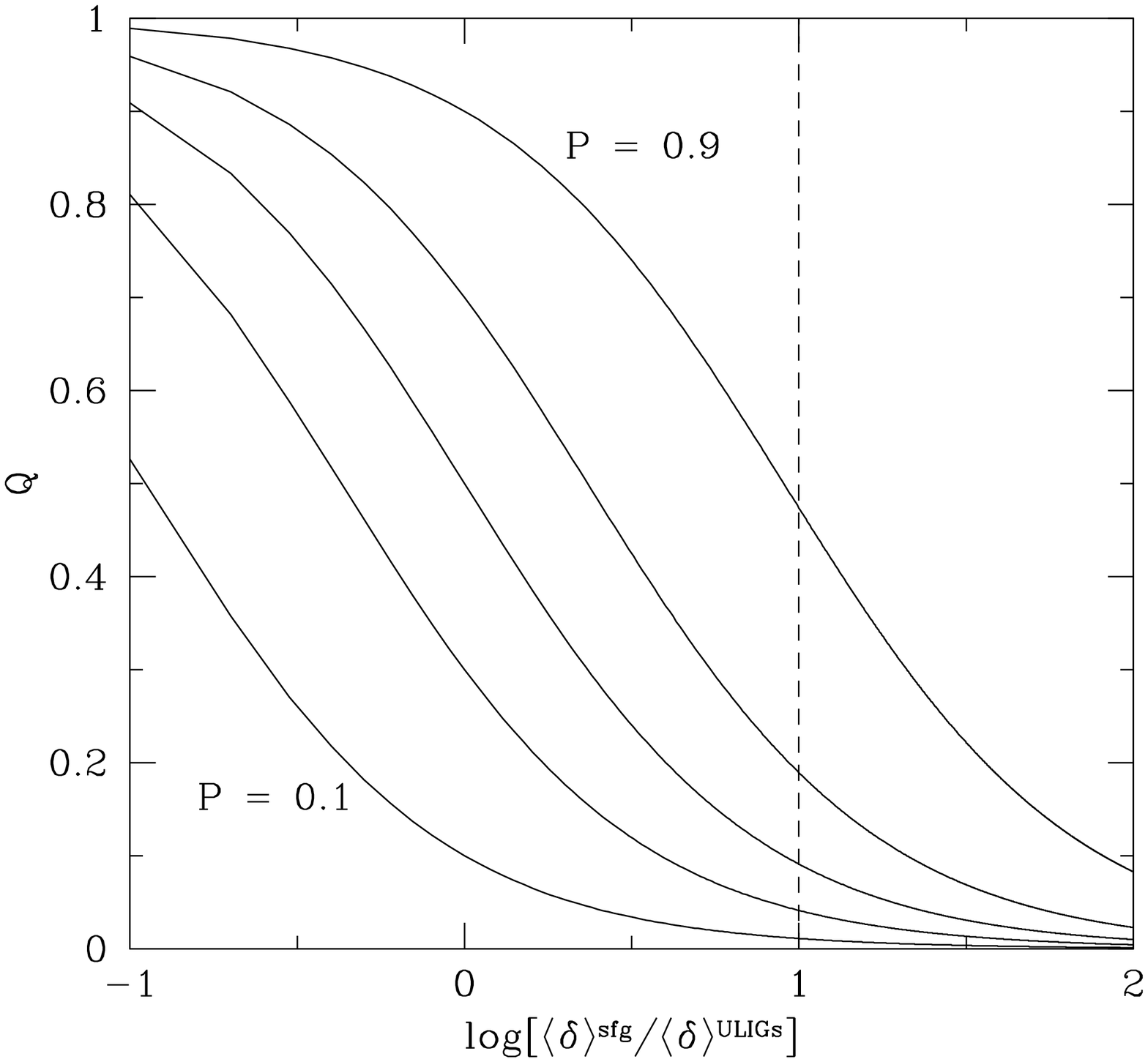}
\caption{Fraction of gas being injected into the IGM that originates in ULIGs, $Q$, as a function of the ratio of the average ejection efficiency of all other star-forming galaxies to that in ULIGs.  The different lines represent different values of $P$, the fractional contribution of ULIGs to the total star formation rate density of the universe at a given $z$, in steps of 0.2. The vertical dotted line indicates $\langle \delta \rangle^{\mathrm{sfg}} / \langle \delta \rangle^{\mathrm{ULIGs}} = 10$, which matches current measurements.  Note that in this case $Q \ga 0.1$ only if $P \ga 0.5$.}
\label{q_v_dd}
\end{figure}

\begin{figure}
\plotone{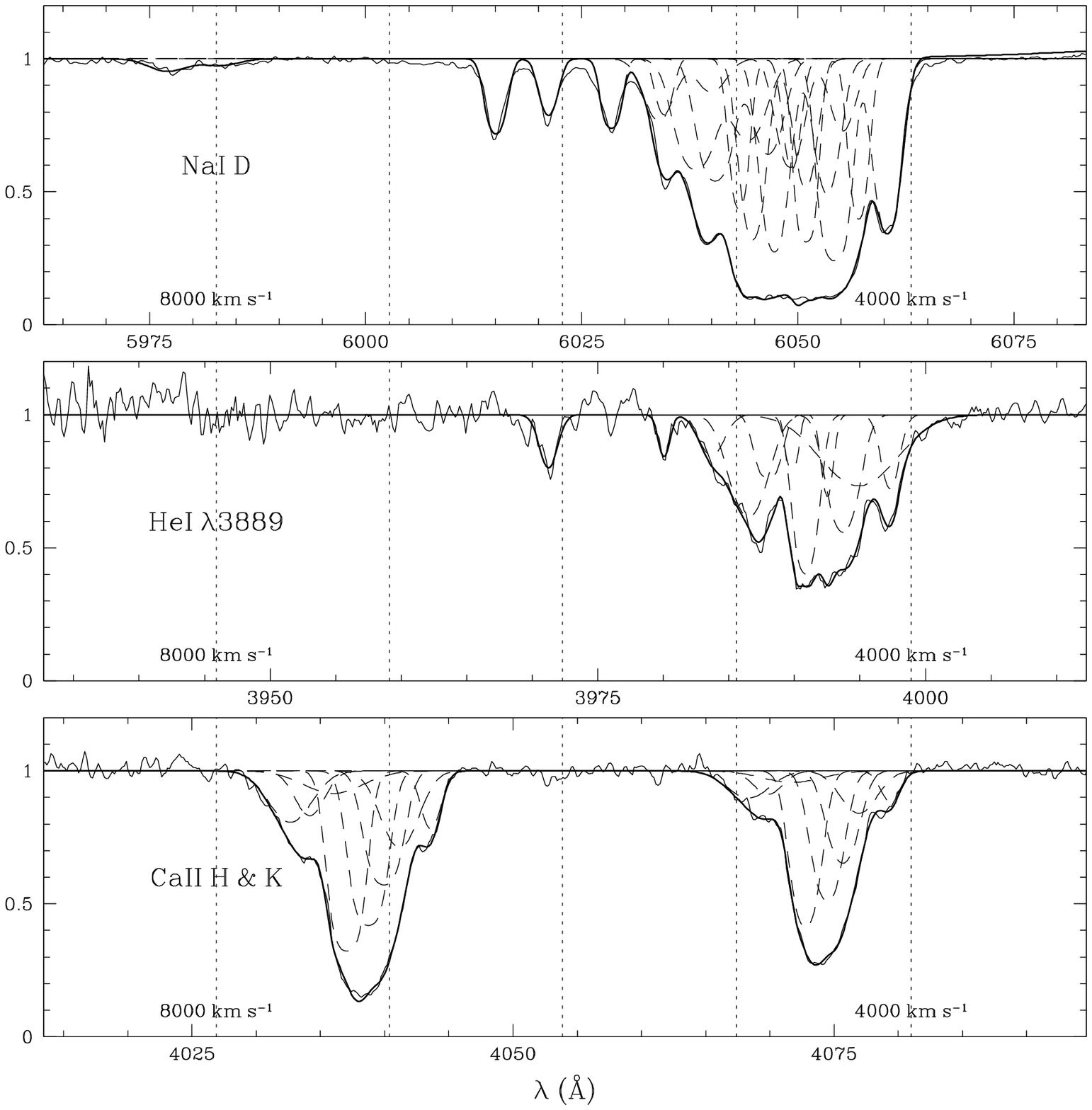}
\caption{Spectra of absorption lines in Mrk 231.  As in Fig. \ref{spectra}, thin solid lines indicate the data, thick solid lines indicate the model, and dashed lines indicate the model components.  The vertical dotted lines are lines of constant blueshifted velocity relative to systemic, spaced at intervals of 1000 \kms, for \ion{Na}{1} D $\lambda 5896$, \ion{He}{1} $\lambda 3889$, and \ion{Ca}{2} H $\lambda3968$.}
\label{mrk231}
\end{figure}

\begin{figure}
\plotone{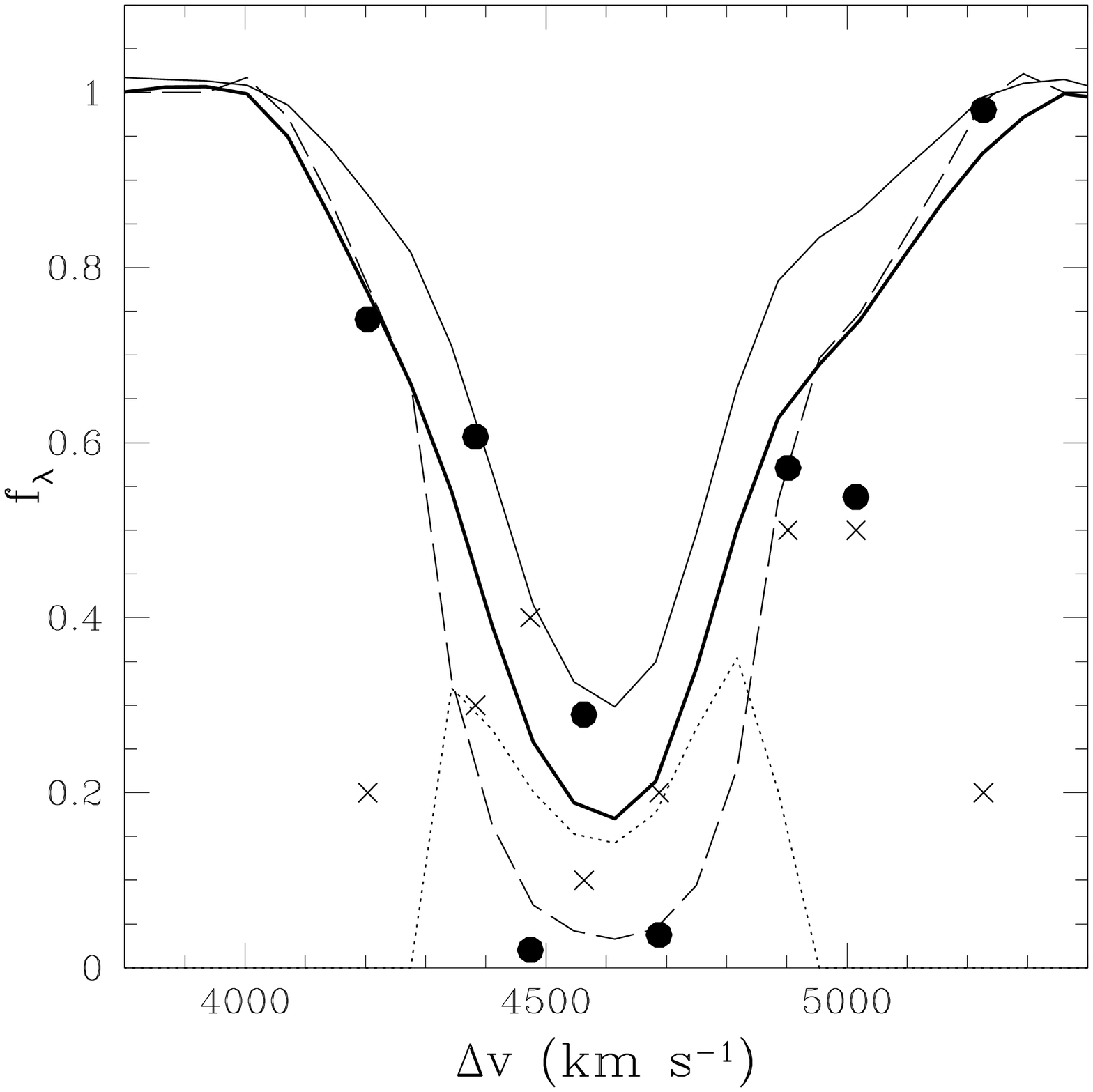}
\caption{\ion{Ca}{2} H \& K lines in Mrk 231, plotted as a function of blueshifted velocity relative to systemic.  The solid lines represent the data, binned in velocity space by 68 \kms\ and boxcar smoothed over 3 bins.  The thin line is \ion{Ca}{2} H $\lambda3968$ and the thick line is \ion{Ca}{2} K $\lambda3934$.  The dotted line is $(1 - C_f)$, where $C_f$ is the covering fraction as a function of velocity.  Finally, the dashed line is $e^{-2 \tau_1}$, where $\tau_1$ is the optical depth of \ion{Ca}{2} H.  The circles and crosses denote the values of $1-C_f$ and $e^{-2 \tau_1}$ that we derive using our curve-of-growth fitting method.  Note that at high optical depths (i.e. near line center) $C_f$ is smaller and the line intensity is largely determined by $C_f$, rather than $\tau$.}
\label{mrk231caii}
\end{figure}

\clearpage

\begin{deluxetable}{llclccccccc}
\rotate
\tabletypesize{\scriptsize}
\tablecaption{Sample Properties \label{sprop}}
\tablewidth{0pt}
\tablehead{
\colhead{IRAS FSC} & \colhead{$z$} & \colhead{log($\frac{L_{\mathrm{IR}}}{L_{\sun}}$)} & \colhead{Type} & \colhead{$M_{K^{\prime}}$} & \colhead{$M_R$} &\colhead{$r_e$} &\colhead{$\mu_e$} &\colhead{Date} & \colhead{Seeing} & \colhead{Exp. Time} \\
\colhead{(1)} & \colhead{(2)} & \colhead{(3)} & \colhead{(4)} & \colhead{(5)} & \colhead{(6)} & \colhead{(7)} & \colhead{(8)} & \colhead{(9)} & \colhead{(10)} & \colhead{(11)}
}
\startdata
03250+1606& 0.1289&  12.06&L&  -25.40&  -21.98&    5.1&   21.2&  2&  0.9&2$\times$900\\
04313$-$1649& 0.2672&  12.55&L&  -24.54&  -22.01&    6.2&   21.8&  3&  0.8&2$\times$1200\\
05024$-$1941& 0.1935&  12.43&S2&  -25.68&  -22.42&    4.7&   20.8&  2&  0.9&2$\times$900\\
05189$-$2524& 0.04274&  12.07&S2&  -25.91&  -21.48&    4.3&   21.3&  3&  0.8&900\\
08559+1053& 0.1481&  12.16&S2&  -26.14&  -22.51&    6.2&   21.5&  1&  1.1&1200\\
09039+0503& 0.1254&  12.07&L&  -24.75&  -21.35&\nodata&\nodata&  2&  0.9&2$\times$900\\
09116+0334& 0.1454&  12.11&L&  -25.96&  -22.11&    4.1&   20.7&  1&  1.1&1200\\
09539+0857& 0.1290&  12.03&L&  -23.94&  -20.54&    3.1&   20.9&  2&  0.9&2$\times$900\\
10190+1322:W& 0.0766&  12.00\tablenotemark{a}&H&  -24.93\tablenotemark{a}&  -21.69\tablenotemark{a}&\nodata&\nodata&  2&  0.9&900\\
10190+1322:E& 0.0759&&L&  -25.30\tablenotemark{a}&  -21.09\tablenotemark{a}&\nodata&\nodata&  2&  0.9&900\\
10378+1108& 0.1367&  12.26&L&  -24.94&  -22.15&    6.1&   21.8&  3&  0.8&2$\times$900\\
11387+4116& 0.1487&  12.18&H&  -25.05&  -21.52&    3.5&   20.5&  3&  0.8&2$\times$900\\
\tableline
12540+5708\tablenotemark{b}& 0.04217&  12.50&S1&  -27.08&  -22.52&    3.3&   19.8&  3&  0.8&300\\
\enddata
\tablecomments{Col.(1): Target name.  FSC $=$ Faint Source Catalog.  Col.(2): Redshift; see Table \ref{z}.  Col.(3): Infrared luminosity, computed from a single-temperature dust-emissivity fit to the fluxes in all four IRAS bands; see \citealt{per87} or \citealt{sm96}.  Col.(4): Optical spectral type.  H = \ion{H}{2}-region-like, L = LINER, S1 = Seyfert 1, S2 = Seyfert 2.  From \citealt*{vks99a}, except for F$04313-1649$ and F$10190+1322$, for which we used the present dataset.  Col.(5): $K^{\prime}$-band absolute magnitude, from \citealt*{vks02}.  Col.(6): $R$-band absolute magnitude, from \citealt{vks02}.  Col.(7): Effective radius from $R$-band surface brightness profile, in kpc.  From \citealt{vks02}.  Col.(8): Surface brightness inside $r_e$, in mag arcs$^{-2}$.  From \citealt{vks02}.  Col.(9): UT date of observation.  $1=$ 2001 Jan 24, $2=$ 2001 Feb 27, $3=$ 2001 Feb 28.  Col.(10): Average seeing on the date of observation, in arcseconds.  Col.(11): Exposure time, in seconds.}
\tablenotetext{a}{F$10190+1322$ has two nuclei separated by 6.2 kpc.  $L_{\mathrm{IR}}$ refers to the object as a whole, i.e. both nuclei at once.  The values of $M_{K^{\prime}}$ and $M_R$ for each nucleus are scaled from the values for the object as a whole ($M_{K^{\prime}} = -25.45$ and $M_R = -21.79$) using the absolute magnitudes within 2 kpc of each nucleus.
}
\tablenotetext{b}{Mrk 231}
\end{deluxetable}

\begin{deluxetable}{lcccrcccr}
\tablecaption{Measured Properties of \ion{Na}{1} D Lines\label{mprop}}
\tablewidth{0pt}
\tablehead{
\colhead{IRAS FSC} & \colhead{No.} & \colhead{$\lambda_1$} & \colhead{$\tau_{1,c}$} & \colhead{$b$} & \colhead{$C_f$} & \colhead{$R$} & \colhead{$W_{\mathrm{eq}}^{\mathrm{rest}}$} & \colhead{$\%$} \\
\colhead{(1)} & \colhead{(2)} & \colhead{(3)} & \colhead{(4)} & \colhead{(5)} & \colhead{(6)} & \colhead{(7)} & \colhead{(8)} & \colhead{(9)}
}
\startdata
03250+1606&  1&  6654.7& 0.22&   353& 1.00& 1.86&  6.7&    14\\
04313$-$1649&  1&  7473.2& 0.58&   124& 0.44& 1.68&  2.5&    15\\
05024$-$1941&  1&  7002.7& 1.72&   165& 0.25& 1.38& 9.7&     4\\
&  2&  7037.0& 3.14&    27& 0.40&1.24&&\\
&  3&  7038.9& 0.17&   333& 1.00&1.89&&\\
05189$-$2524&  1&  6139.4& 0.73&   419& 0.31& 1.62&  5.5&    12\\
&  2&  6148.0& 3.75&    85& 0.16&1.21&&\\
08559+1053&  1&  6772.2& 0.27&   313& 0.56& 1.83&  3.9&    19\\
09039+0503&  1&  6629.8& 0.42&   390& 0.26& 1.75&  3.5&    28\\
&  2&  6639.7& 0.63&    50& 0.24&1.66&&\\
09116+0334&  1&  6752.4& 0.18&   338& 0.72& 1.88&  3.9&    18\\
09539+0857&  1&  6652.0& 0.36&   282& 0.62& 1.78&  4.9&    16\\
10190+1322:W&  1&  6349.0& 1.20&   114& 0.68& 1.48&  6.1&    15\\
10190+1322:E&  1&  6346.4& 0.92&   200& 0.61& 1.56&  9.1&     10\\
&  2&  6352.0& 0.41&    94& 1.00&1.76&&\\
10378+1108&  1&  6683.8& 0.06&   750& 0.77& 1.96&  7.3&    11\\
&  2&  6693.1& 0.17&   479& 0.60&1.89&&\\
11387+4116&  1&  6765.2& 0.19&   117& 1.00& 1.88&  8.0&     14\\
&  2&  6772.2& 0.74&   194& 0.67&1.62&&\\
\enddata
\tablecomments{Col.(1): Target name.  Col.(2): Component number.  Components listed in order from highest to lowest velocity relative to systemic.  Col.(3): Centroid of D$_1$ $\lambda 5896$ line, in \AA.  Col.(4): Central optical depth of D$_1$ line.  Optical depth of D$_2$ $\lambda 5890$ is twice this value.  Col.(5): Doppler parameter of absorbing gas in \kms, equal to $\mathrm{FWHM}/(2 \sqrt{\ln 2})$.  Before conversion to $b$, the FWHM is corrected in quadrature for a 65 \kms\ instrumental resolution.  Col.(6): Covering fraction of absorbing gas.  Col.(7): Doublet equivalent width ratio $R = W_\mathrm{eq}^{\lambda5890} / W_\mathrm{eq}^{\lambda5896}$.  Col.(8): Total rest-frame equivalent width of \ion{Na}{1} D absorption complex, in \AA.  The rest-frame equivalent width is related to the observed equivalent width by $W_{\mathrm{eq}}^{\mathrm{rest}} = W_{\mathrm{eq}}^{\mathrm{obs}} / (1 + z)$.  Col.(9): Percent stellar contribution to \ion{Na}{1} D line, estimated from the equivalent width of the \ion{Mg}{1} b triplet.}
\end{deluxetable}

\begin{deluxetable}{lccccccccccc}
\rotate
\tablecaption{Absorption Line Equivalent Widths\label{alew}}
\tablewidth{0pt}
\tablehead{
\colhead{} & \colhead{\ion{Mg}{1} b} & \colhead{H$\gamma$} & \colhead{H$\delta$} & \colhead{H8} & \colhead{H9} & \colhead{H10} & \colhead{H11} & \colhead{\ion{Ca}{2} K} & \colhead{\ion{Ca}{2}} & \colhead{\ion{Ca}{2}} & \colhead{\ion{Ca}{2}}\\
\colhead{IRAS FSC} & \colhead{} & \colhead{4340} & \colhead{4102} & \colhead{3889} & \colhead{3835} & \colhead{3798} & \colhead{3771} & \colhead{3934} & \colhead{8498} & \colhead{8542} & \colhead{8662}\\
\colhead{(1)} & \colhead{(2)} & \colhead{(3)} & \colhead{(4)} & \colhead{(5)} & \colhead{(6)} & \colhead{(7)} & \colhead{(8)} & \colhead{(9)} & \colhead{(10)} & \colhead{(11)} & \colhead{(12)}
}
\startdata
03250+1606&    1.9&\nodata&\nodata&   10.4&    \phn8.9&    \phm{::}9.0::&\nodata&    5.7&\nodata&\nodata&    \phm{::}2.9::\\
04313$-$1649&    0.8&\nodata&\nodata&    \phn\phm{::}6.2::&    \phn7.8&    5.5&    \phm{::}5.3::&    3.2&\nodata&\nodata&\nodata\\
05024$-$1941&    0.7&    3.3&\nodata&    \phn8.6&    \phn7.8&    4.7&\nodata&    3.2&\nodata&\nodata&\nodata\\
05189$-$2524&    1.3&    4.1&    5.7&    \phn5.7&    \phn6.4&    6.0&\nodata&    1.9&    1.1&    2.0&    2.0\\
08559+1053&    1.4&\nodata&\nodata&\nodata&\nodata&    3.2&    3.9&\nodata&\nodata&    \phm{::}2.2::&\nodata\\
09039+0503&    2.0&\nodata&\nodata&\nodata&    \phn\phm{::}6.2::&    5.9&    \phm{::}2.9::&    3.1&\nodata&\nodata&\nodata\\
09116+0334&    1.4&\nodata&    7.9&    \phn7.9&    \phn7.4&    5.2&\nodata&\nodata&    1.2&    2.2&\nodata\\
09539+0857&    1.6&\nodata&\nodata&    \phn9.1&    \phn6.6&    4.5&\nodata&    3.4&\nodata&    \phm{::}2.6::&    2.1\\
10190+1322:W&    1.8&\nodata&\nodata&    \phn\phm{::}7.4::&    \phn6.9&    \phm{::}4.7::&\nodata&    3.5&    \phm{::}2.3::&    2.8&\nodata\\
10190+1322:E&    1.8&\nodata&\nodata&   \phm{::}15.5::&   \phm{::}17.5::&\nodata&\nodata&    4.5&\nodata&    3.0&\nodata\\
10378+1108&    1.5&\nodata&\nodata&\nodata&    \phn7.1&    5.6&\nodata&    5.2&    \phm{::}1.6::&    \phm{::}3.8::&    1.5\\
11387+4116&    2.2&\nodata&    6.2&   10.4&    \phn7.3&    5.3&\nodata&\nodata&    1.8&    4.0&    \phm{::}3.6::\\
\enddata
\tablecomments{Col.(1): Target name.  Col.(2)-(12): Equivalent widths of selected absorption lines, in \AA.  Uncertainties are generally $\sim20\%$; numbers marked with a (::) have an uncertainty of $\sim40\%$.  Values quoted are rest-frame equivalent widths: $W_{\mathrm{eq}}^{\mathrm{rest}} = W_{\mathrm{eq}}^{\mathrm{obs}}/(1+z)$.  Wavelengths listed in the table header are wavelengths in air.}
\end{deluxetable}

\begin{deluxetable}{lllll}
\tabletypesize{\scriptsize}
\tablecaption{Redshifts \label{z}}
\tablewidth{0pt}
\tablehead{
\colhead{IRAS FSC} & \colhead{$z_\mathrm{abs}$} & \colhead{$z_\mathrm{em}$} & \colhead{$z_\mathrm{other}$} & \colhead{$z_\mathrm{adopt}$} \\
\colhead{(1)} & \colhead{(2)} & \colhead{(3)} & \colhead{(4)} & \colhead{(5)}
}
\startdata
03250+1606& 0.1290& 0.1289& \nodata & 0.1289\\
04313$-$1649& 0.2672& 0.2675&\nodata& 0.2672\\
05024$-$1941& 0.1936& 0.1935& 0.1914\tablenotemark{a}& 0.1935\\
05189$-$2524& 0.0429& 0.0427& 0.04274\tablenotemark{b}& 0.04274\\
08559+1053& 0.1482& 0.1480&\nodata& 0.1481\\
09039+0503& 0.1254& 0.1250&\nodata& 0.1254\\
09116+0334& 0.1454& 0.1453&\nodata& 0.1454\\
09539+0857& 0.1290& 0.1291&\nodata& 0.1290\\
10190+1322:W& 0.0767& 0.0766&\nodata& 0.0766\\
10190+1322:E& 0.0758& 0.0760&\nodata& 0.0759\\
10378+1108& 0.1367& 0.1363&\nodata& 0.1367\\
11387+4116& 0.1487& 0.1487&\nodata& 0.1487\\
\tableline
12540+5708&\nodata&\nodata& 0.04217\tablenotemark{c}& 0.04217\\
\enddata
\tablecomments{
Col.(1): Target name.  Col.(2): Average redshift measured from stellar absorption lines, including as many as possible of the following lines: the Balmer lines $\mathrm{H}\gamma-\mathrm{H}10$, \ion{Ca}{2} K, and the \ion{Ca}{2} triplet.  Col.(3): Average redshift measured from emission lines, including as many as possible of the following lines: the Balmer lines $\mathrm{H}\alpha-\mathrm{H}\delta$, [\ion{N}{2}] $\lambda\lambda6548,6583$, [\ion{O}{3}] $\lambda\lambda4959,5007$, [\ion{O}{1}] $\lambda\lambda6300,6363$, [\ion{S}{2}] $\lambda\lambda6716,6731$.  (We also used \ion{He}{2} $\lambda4686$ and [\ion{Ni}{2}] $\lambda7378$ in a couple of objects.)  Col.(4): Redshifts from other sources: CO, \ion{H}{1}, and higher-excitation emission lines.  Col.(5): Adopted redshift.  CO and \ion{H}{1} data are used where possible.  In other cases, where the emission- and absorption-line redshifts differ by no more than $\Delta z = \pm 0.0002$, we take the average of the two.  Otherwise, we use the absorption-line redshift.
}
\tablenotetext{a}{Redshift measured from [\ion{O}{3}] $\lambda\lambda4959,5007$, using this dataset.}
\tablenotetext{b}{Redshift measured from CO emission; see \citealt*{sss91}.}
\tablenotetext{c}{Redshift measured from \ion{H}{1} absorption; see \citealt*{cwu98}.}
\end{deluxetable}

\begin{deluxetable}{lcllrrlrcclll}
\tabletypesize{\scriptsize}
\tablecaption{Computed Properties\label{cprop}}
\tablewidth{0pt}
\tablehead{
\colhead{IRAS FSC} & \colhead{No.} & \colhead{$N$(\ion{Na}{1})} & \colhead{$N$(H)} & \colhead{$\Delta v$} & \colhead{$\Delta V$} & \colhead{$\dot{M}$} & \colhead{$\dot{M}_{\mathrm{tot}}$} & \colhead{$\frac{\mathrm{SFR}}{\alpha}$} & \colhead{$\alpha$} & \colhead{$\frac{\dot{M}_{\mathrm{tot}}}{\mathrm{SFR}}$} & \colhead{$\frac{\dot{M}_\mathrm{esc}}{\dot{M}_{\mathrm{tot}}}$} & \colhead{$\frac{\dot{M}_\mathrm{esc}}{\mathrm{SFR}}$}\\
\colhead{(1)} & \colhead{(2)} & \colhead{(3)} & \colhead{(4)} & \colhead{(5)} & \colhead{(6)} & \colhead{(7)} & \colhead{(8)} & \colhead{(9)} & \colhead{(10)}
& \colhead{(11)} & \colhead{(12)} & \colhead{(13)}
}
\startdata
03250+1606&  1&    \phn2.8&    1.6&    137&    431&  \phn22&    22&   198&  0.8&  0.14&      0.01&      0.00\\
04313$-$1649&  1&    \phn2.6&    1.5&      9&    112&   \phn\phn0.0&     0&   612&  0.8&  0.00&      0.00&      0.00\\
05024$-$1941&  1&   10&    3.0&   1538&   1676& 117&   122&   464&  0.6&  0.44&      0.96&      0.42\\
&  2&    \phn3.0&    1.7&     73&     96&   \phn\phn5.0&&&&&&\\
&  3&    \phn2.0&    1.4&     -9&    268&   \phn\phn0.0&&&&&&\\
05189$-$2524&  1&   11&    3.1&    501&    849&  \phn49&    53&   203&  0.4&  0.66&      0.31&      0.21\\
&  2&   11&    3.1&     81&    152&   \phn\phn4.2&&&&&&\\
08559+1053&  1&    \phn3.0&    1.7&    -49&    212&   \phn\phn0.0&     0&   249&  0.6&  0.00&      0.00&      0.00\\
09039+0503&  1&    \phn5.8&    2.3&    331&    656&  \phn20&    20&   203&  0.8&  0.12&      0.28&      0.03\\
&  2&    \phn1.1&    1.0&   -118&    -77&   \phn\phn0.0&&&&&&\\
09116+0334&  1&    \phn2.2&    1.4&    120&    402&  \phn13&    13&   222&  0.8&  0.07&      0.00&      0.00\\
09539+0857&  1&    \phn3.6&    1.8&    284&    519&  \phn33&    33&   185&  0.8&  0.22&      0.09&      0.02\\
10190+1322:W&  1&    \phn4.9&    2.1&     14&    108&   \phn\phn0.0&     0\tablenotemark{a}&   172\tablenotemark{a}&  0.8\tablenotemark{a}&  0.00\tablenotemark{a}&      0.00\tablenotemark{a}&      0.00\tablenotemark{a}\\
10190+1322:E&  1&    \phn6.5&    2.4&    -58&    109&   \phn\phn0.0&&&&&&\\
&  2&    \phn1.4&    1.2&   -321&   -243&   \phn\phn0.0&&&&&&\\
10378+1108&  1&    \phn1.5&    1.2&    892&   1517&  \phn85&   133&   314&  0.8&  0.53&      0.58&      0.31\\
&  2&    \phn2.9&    1.6&    476&    875&  \phn48&&&&&&\\
11387+4116&  1&    \phn0.80&    0.89&    413&    511&  \phn38&    53&   261&  0.8&  0.26&      0.00&      0.00\\
&  2&    \phn5.1&    2.1&    104&    265&  \phn15&&&&&&\\
\enddata
\tablecomments{Col.(1): Target name.  Col.(2): Component number.  Col.(3): Column density of \ion{Na}{1}, in units of $10^{13}$ cm$^{-2}$.  Col.(4): Column density of H, in units of $10^{21}$ cm$^{-2}$.  Col.(5): Velocity of the component relative to systemic, in \kms: $\Delta v = v_{\mathrm{sys}} - v$.  Blueshifted velocities are positive.  Col.(6): ``Maximum'' blueshifted velocity of the component, in \kms: $\Delta V = \Delta v + \onehalf \mathrm{FWHM}$.  Col.(7): Mass outflow rate for the component in \smpy.  Col.(8): Total mass outflow rate for the object in \smpy.  Col.(9): Star formation rate in \smpy, derived from $L_{\mathrm{IR}}$ using the prescription in \citealt{kenn98} and parameterized by the fractional contribution of star formation to the infrared luminosity, which is given by $\alpha$ ($0 \leq \alpha \leq 1$).  Col.(10): Estimated fraction of infrared (or $\sim$ bolometric) luminosity powered by star formation.  Col.(11): Reheating efficiency $\eta$, equal to the total mass outflow rate normalized by the corresponding global star formation rate, and including the proper values for $\alpha$.  Col.(12): Escape fraction $f_\mathrm{esc}$, equal to the mass outflow rate of escaping gas divided by the total mass outflow rate.  Col.(13): Ejection efficiency $\delta$, equal to the total mass outflow rate of escaping gas normalized by the corresponding global star formation rate.}
\tablenotetext{a}{The infrared luminosity $L_{\mathrm{IR}}$ for F$10190+1322$ refers to both nuclei, so these quantities also refer to the object as a whole.}
\end{deluxetable}

\begin{deluxetable}{clrcr}
\tabletypesize{\scriptsize}
\tablecaption{Na to H Conversion Factors \label{natoh}}
\tablewidth{0pt}
\tablehead{
\colhead{Ref.} & \colhead{$\beta$} & \colhead{$\gamma$} & \colhead{H$_2$} & \colhead{$\frac{N_2(\mathrm{H})}{N_1(\mathrm{H})}$}\\
\colhead{(1)} & \colhead{(2)} & \colhead{(3)} & \colhead{(4)} & \colhead{(5)}
}
\startdata
1   & $2.3\pm0.4$   & \nodata & n       & \nodata \\
2   & 2.3           & -35.1   & n       & $5-24$  \\
3   & 2.1           & \nodata & b       & \nodata \\
4   & 2.11          & -31.3   & y       & $6-29$  \\
5   & $1.04\pm0.08$ & -9.09   & y       & $36-41$ \\
6   & $1.61\pm0.17$ & \nodata & n       & \nodata \\
5,6\tablenotemark{a} & $1.62\pm0.13$ & \nodata & y       & \nodata \\
\enddata
\tablerefs{
(1) \citealt{hobbs74a}; (2) \citealt{hobbs74b}; (3) \citealt{hobbs76}; (4) \citealt{stokes78}; (5) \citealt*{fvg85}; (6) \citealt{herbig93}.
}
\tablecomments{Col.(1): Reference.  Col. (2): Slope of conversion equation: $\log[N$(\ion{Na}{1})$] = \beta \log[N$(H)$] + \gamma$.  Col. (3): Intercept of conversion equation.  Col. (4): Is molecular hydrogen included in the calculation of the relation?  y = yes [i.e. $N(\mathrm{H}) = N($\ion{H}{1}$) + 2 N(\mathrm{H}_2)$]; n = no [i.e. $N(\mathrm{H}) = N($\ion{H}{1}$)$]; b = both (i.e. both kinds of data included).  Col. (5): The ratio between $N$(H) determined using the conversion relationships in this table (``method 2'') and $N$(H) determined using the depletion correction from \citealt{ss96} (-0.95) but with no ionization correction (``method 1'').  The range of values is for the data in our sample.  Note that we are able to compute $N_2(\mathrm{H})$ for only those sources that include $\gamma$.}
\tablenotetext{a}{\citealt{herbig93} re-fitted the data of \citealt{fvg85}, excluding cases where $N$(H) along a given sightline was subdivided into individual kinematic components and using values for $N$(\ion{Na}{1}) from the earlier work of Hobbs where applicable.}
\end{deluxetable}

\begin{deluxetable}{lcccccc}
\tablecaption{Host Galaxy Kinematics\label{kin}}
\tablewidth{0pt}
\tablehead{
\colhead{IRAS FSC} & \colhead{\onehalf FWHM$_\mathrm{avg}^\mathrm{em}$} & \colhead{\onehalf FWHM$_\mathrm{avg}^\mathrm{abs}$} & \colhead{$\sigma_0$} & \colhead{\onehalf $v_\mathrm{pp}$} & \colhead{$v_\mathrm{rot}$} & \colhead{$v_\mathrm{esc}$} \\
\colhead{(1)} & \colhead{(2)} & \colhead{(3)} & \colhead{(4)} & \colhead{(5)} & \colhead{(6)} & \colhead{(7)}
}
\startdata
03250+1606&    272&    256&    199&    198&\nodata&    724\\
04313$-$1649&    137&\nodata&    154&\nodata&\nodata&    559\\
05024$-$1941&    272&\nodata&    248&\nodata&\nodata&    902\\
05189$-$2524&    242&    198&    156&\nodata&\nodata&    621\\
08559+1053&    223&    222&    193&    157&\nodata&    702\\
09039+0503&    190&\nodata&\nodata&\nodata&\nodata&    488\\
09116+0334&    274&    197&    234&\nodata&\nodata&    849\\
09539+0857&    138&    140&    152&\nodata&\nodata&    552\\
10190+1322:W&    122&    130&\nodata&    122&    245\tablenotemark{a}&    630\\
10190+1322:E&    198&    242&\nodata&\nodata&    175&    621\\
10378+1108&    228&    231&    151&\nodata&\nodata&    594\\
11387+4116&    185&    215&    232&\nodata&\nodata&    843\\
\tableline
12540+5708&\nodata&\nodata&    371&\nodata&\nodata&\nodata\\
\enddata
\tablecomments{Col.(1): Target name.  Col.(2): One-half the average FWHM of emission lines.  Col.(3): One-half the average FWHM of \ion{Ca}{2} triplet lines.  Col.(4): Central velocity dispersion, computed from the $R$-band fundamental plane.  Col.(5): One-half the peak-to-peak velocity amplitude determined from position-velocity diagrams.  Col.(6): Rotation curve amplitudes from the Pa$\alpha$ line; from \citealt{murphy01}.  Col.(7): Escape velocity, computed using a singular isothermal sphere with $r_\mathrm{max}/r$ = 10 and the maximum of \onehalf FWHM$_\mathrm{avg}^\mathrm{em}$, \onehalf FWHM$_\mathrm{avg}^\mathrm{CaII}$, $\sqrt{2}\sigma_0$, the rotation speed measured in position-velocity space, and, in the case of F$10190+1322$, the measured rotation speeds. }
\tablenotetext{a}{Corrected for a galaxy inclination of $40-50\degr$.}
\end{deluxetable}

\begin{deluxetable}{lccccccccc}
\tablecaption{Emission Line Widths \label{elw}}
\tablewidth{0pt}
\tablehead{
\colhead{} & \colhead{[\ion{O}{3}]} & \colhead{[\ion{O}{3}]} & \colhead{[\ion{O}{1}]} & \colhead{[\ion{N}{2}]} & \colhead{H$\alpha$} & \colhead{[\ion{N}{2}]} & \colhead{[\ion{S}{2}]} & \colhead{[\ion{S}{2}]} & \colhead{}\\
\colhead{IRAS FSC} & \colhead{4959} & \colhead{5007} & \colhead{6300} & \colhead{6548} & \colhead{6563} & \colhead{6583} & \colhead{6716} & \colhead{6731} & \colhead{avg}\\
\colhead{(1)} & \colhead{(2)} & \colhead{(3)} & \colhead{(4)} & \colhead{(5)} & \colhead{(6)} & \colhead{(7)} & \colhead{(8)} & \colhead{(9)} & \colhead{(10)}
}
\startdata
03250+1606&      \nodata&    \phm{:}548:&    \phm{:}484:&    549&    622&    540&    \phm{:}522:&      \nodata&    544\\
04313$-$1649&    \phm{:}294:&    \phm{:}311:&    \phm{:}284:&      \nodata&    257&      \nodata&    254&    244&    274\\
05024$-$1941&      \nodata&   \phm{:}1167:\phn&    \phm{:}379:&      \nodata&    \phm{:}682:&    \phm{:}518:&    \phm{:}240:&   \phm{:}280:&    544\\
05189$-$2524&    768&    821&    260&      \nodata&    441&      \nodata&    299&    312&    483\\
08559+1053&    521&    324&    552&      \nodata&    356&    392&    \phm{:}489:&    488&    446\\
09039+0503&      \nodata&    371&    424&      \nodata&    302&    366&    371&    443&    380\\
09116+0334&      \nodata&    \phm{:}684:&    \phm{:}483:&    563&    523&    525&      \nodata&    511&    548\\
09539+0857&    \phm{:}258:&    251&    253&    341&    255&    291&      \nodata&      \nodata&    275\\
10190+1322:W&      \nodata&    \phm{:}255:&    \phm{:}289:&    287&    184&    212&    \phm{:}245:&    247&    245\\
10190+1322:E&      \nodata&      \nodata&    \phm{:}362:&    317&    507&    399&      \nodata&      \nodata&    397\\
10387+1108&    510&    432&    467&      \nodata&      \nodata&    416&      \nodata&      \nodata&    456\\
11387+4116&      \nodata&    305&    \phm{:}405:&    506&    235&    312&    418&    412&    370\\
\enddata
\tablecomments{Col.(1): Target name.  Col.(2)-(9): Full-widths at half-maximum of selected emission lines, in \kms.  Uncertainties are generally $5-10\%$; numbers marked with a (:) have an uncertainty of $20\%$.  Values have been corrected in quadrature for a finite instrumental resolution of 65 \kms.  Wavelengths listed in the table header are wavelengths in air.  Col.(10): Average FWHM.}
\end{deluxetable}

\begin{deluxetable}{lcccccccccccc}
\rotate
\tablecaption{Absorption Line Widths\label{alw}}
\tablewidth{0pt}
\tablehead{
\colhead{} & \colhead{H$\gamma$} & \colhead{H$\delta$} & \colhead{H8} & \colhead{H9} & \colhead{H10} & \colhead{H11} & \colhead{H} & \colhead{\ion{Ca}{2} K} & \colhead{\ion{Ca}{2}} & \colhead{\ion{Ca}{2}} & \colhead{\ion{Ca}{2}} & \colhead{\ion{Ca}{2} T}\\
\colhead{IRAS FSC} & \colhead{4340} & \colhead{4102} & \colhead{3889} & \colhead{3835} & \colhead{3798} & \colhead{3771} & \colhead{avg} & \colhead{3934}  & \colhead{8498} & \colhead{8542} & \colhead{8662} & \colhead{avg}\\
\colhead{(1)} & \colhead{(2)} & \colhead{(3)} & \colhead{(4)} & \colhead{(5)} & \colhead{(6)} & \colhead{(7)} & \colhead{(8)} & \colhead{(9)} & \colhead{(10)} & \colhead{(11)} & \colhead{(12)} & \colhead{(13)}
}
\startdata
03250+1606&      \nodata&      \nodata&   1807\phn&   1474\phn&   \phm{:}\phm{:}1187::\phn&      \nodata&  1489\phn&   1177\phn&      \nodata&      \nodata&    \phm{:}\phm{:}513::&   513\\
04313$-$1649&      \nodata&      \nodata&   \phm{:}\phm{:}1242::\phn&   1200\phn&    906&   \phm{:}\phm{:}1051::\phn&  1100\phn&    490&      \nodata&      \nodata&      \nodata&     \nodata\\
05024$-$1941&    865&      \nodata&   1857\phn&   1277\phn&    975&      \nodata&  1243\phn&    253&      \nodata&      \nodata&      \nodata&     \nodata\\
05189$-$2524&    763&    819&    876&    920&    873&      \nodata&   850&    371&    339&    420&    430&   396\\
08559+1053&      \nodata&      \nodata&      \nodata&      \nodata&    959&    941&   950&      \nodata&      \nodata&    443&      \nodata&   443\\
09039+0503&      \nodata&      \nodata&      \nodata&   \phm{:}\phm{:}1289::\phn&   1216\phn&    \phm{:}\phm{:}664::&  1056\phn&    819&      \nodata&      \nodata&      \nodata&     \nodata\\
09116+0334&      \nodata&   1354\phn&   1404\phn&   1224\phn&    991&      \nodata&  1243\phn&      \nodata&    347&    \phm{:}\phm{:}441::&      \nodata&   394\\
09539+0857&      \nodata&      \nodata&   1860\phn&   1181\phn&   1088\phn&      \nodata&  1376\phn&   1018\phn&      \nodata&    \phm{:}\phm{:}260::&    299&   280\\
10190+1322:W&      \nodata&      \nodata&  \phm{:}\phm{:}1589::\phn&   1072\phn&    \phm{:}\phm{:}927::&      \nodata&  1196\phn&    472&    \phm{:}\phm{:}224::&    297&      \nodata&   260\\
10190+1322:E&      \nodata&      \nodata&   \phm{:}\phm{:}1336::\phn&   \phm{:}\phm{:}2012::\phn&      \nodata&      \nodata&  1674\phn&    582&      \nodata&    \phm{:}\phm{:}483::&      \nodata&   483\\
10378+1108&      \nodata&      \nodata&      \nodata&   1221\phn&   1069\phn&      \nodata&  1145\phn&    999&    \phm{:}\phm{:}392::&    \phm{:}\phm{:}651::&    343&   462\\
11387+4116&      \nodata&   1461\phn&   1777\phn&   1211\phn&   1011\phn&      \nodata&  1365\phn&      \nodata&    365&    581&    \phm{:}\phm{:}343::&   430\\
\enddata
\tablecomments{Col.(1): Target name.  Col.(2)-(7),(9)-(12): Full-widths at half-minimum of selected absorption lines, in \kms.  Uncertainties are generally $\sim20\%$; numbers marked with a (::) have an uncertainty of $\sim40\%$.  Values have been corrected in quadrature for a finite instrumental resolution of 65 \kms.  Wavelengths listed in the table header are wavelengths in air.  Col.(8): Average FWHM of Balmer lines.  Col.(13): Average FWHM of \ion{Ca}{2} triplet lines.}
\end{deluxetable}

\begin{deluxetable}{cccrllrllrl}
\tablecolumns{11}
\tablecaption{Properties of Absorption Lines in Mrk 231 \label{table_mrk231}}
\tablewidth{0pt}
\tablehead{
\colhead{} & \colhead{} & \multicolumn{3}{c}{\ion{Na}{1} D} & \multicolumn{3}{c}{\ion{He}{1} $\lambda3889$} & \multicolumn{3}{c}{\ion{Ca}{2} H} \\
\cline{3-5} \cline{6-8} \cline{9-11} \\
\colhead{$\Delta v$} & \colhead{$C_f$} & \colhead{$\tau_{1,c}$} & \colhead{$b$} & \colhead{$N$} & \colhead{$\tau_c$} & \colhead{$b$} & \colhead{$N$} & \colhead{$\tau_{\mathrm{H},c}$} & \colhead{$b$} & \colhead{$N$}\\
\colhead{(1)} & \colhead{(2)} & \colhead{(3)} & \colhead{(4)} & \colhead{(5)} & \colhead{(6)} & \colhead{(7)} & \colhead{(8)} & \colhead{(9)} & \colhead{(10)} & \colhead{(11)}
}
\startdata
  8062&   0.2& 0.16&  141&  0.80& 0&    0&   \phn0&  0&    0&  0\\
  6164&   0.3& 1.12&   50&  2.0& 1.15&   42&  13&  0&    0&  0\\
  5500&   0.3& 1.43&   41&  2.1& 0.72&   17&   \phn3.3&  0&    0&  0\\
  5227&   0.8& 0.17&  107&  0.65& 0.23&   85&   \phn5.2&  0.01&  107&  0.00\\
  5015&   0.5& 1.01&  118&  4.3& 1.50&  115&  46&  0.31&  120&  1.9\\
  4902&   0.5& 1.27&   96&  4.3& 0.72&   76&  15&  0.28&   72&  1.0\\
  4766&   0.7& 1.01&   47&  1.7& 0.80&   35&   \phn7.5&  0&    0&  0\\
  4688&   0.8& 0.92&   72&  2.4& 0.95&   56&  14&  1.64&   92&  7.7\\
  4563&   0.9& 0.86&   83&  2.5& 0.59&  131&  20&  0.62&   85&  2.7\\
  4474&   0.6& 0.69&   83&  2.0& 1.76&  135&  63&  1.95&   85&  8.4\\
  4383&   0.7& 1.96&   48&  3.4& 0.19&   46&   \phn2.3&  0.26&  142&  1.9\\
  4204&   0.8& 1.90&   76&  5.2& 0.67&   86&  15&  0.15&   66&  0.52\\
\enddata
\tablecomments{Col.(1): Velocity relative to systemic, in \kms: $\Delta v = v_{\mathrm{sys}} - v$.  Blueshifted velocities are positive.  Col.(2): Covering fraction.  Col.(3): Central optical depth of \ion{Na}{1} D$_1$ $\lambda 5896$ line.  Col.(4): Doppler parameter of \ion{Na}{1} component, in \kms.  Before conversion to $b$, the FWHM is corrected in quadrature for a 65 \kms\ instrumental resolution.  Col.(5): Column density of \ion{Na}{1} component, in units of $10^{13}\;\mathrm{cm}^{-2}$.  Col.(6)-(8): Same as (3)-(5), but for \ion{He}{1} $\lambda 3889$.  Col.(9)-(11): Same as (3)-(5), but for \ion{Ca}{2} H.}
\end{deluxetable}

\end{document}